\documentclass{article}
\usepackage{graphicx} 
\usepackage{amsmath}
\usepackage{amssymb}
\usepackage{hyperref}
\usepackage{geometry}
\usepackage{booktabs} %
\usepackage{amsfonts}
\usepackage{multirow}
\usepackage{amsthm}
\usepackage{enumitem}
\usepackage{dcolumn}
\usepackage{bm}
\usepackage{longtable}
\usepackage{float}
\usepackage{tabularx}   %
\usepackage{epsfig}
\input{epsf}
\usepackage{psfrag}
\usepackage{xcolor}
\usepackage{pdfpages}
\usepackage[normalem]{ulem}
\usepackage{subcaption} %
\usepackage[numbers,sort&compress]{natbib}
\usepackage{authblk}

\newtheorem{definition}{Definition}

\title{~\\[-3cm]\textbf{Iterated Agent for Symbolic Regression}\\[0.3cm]
\includegraphics[width=0.2\linewidth]{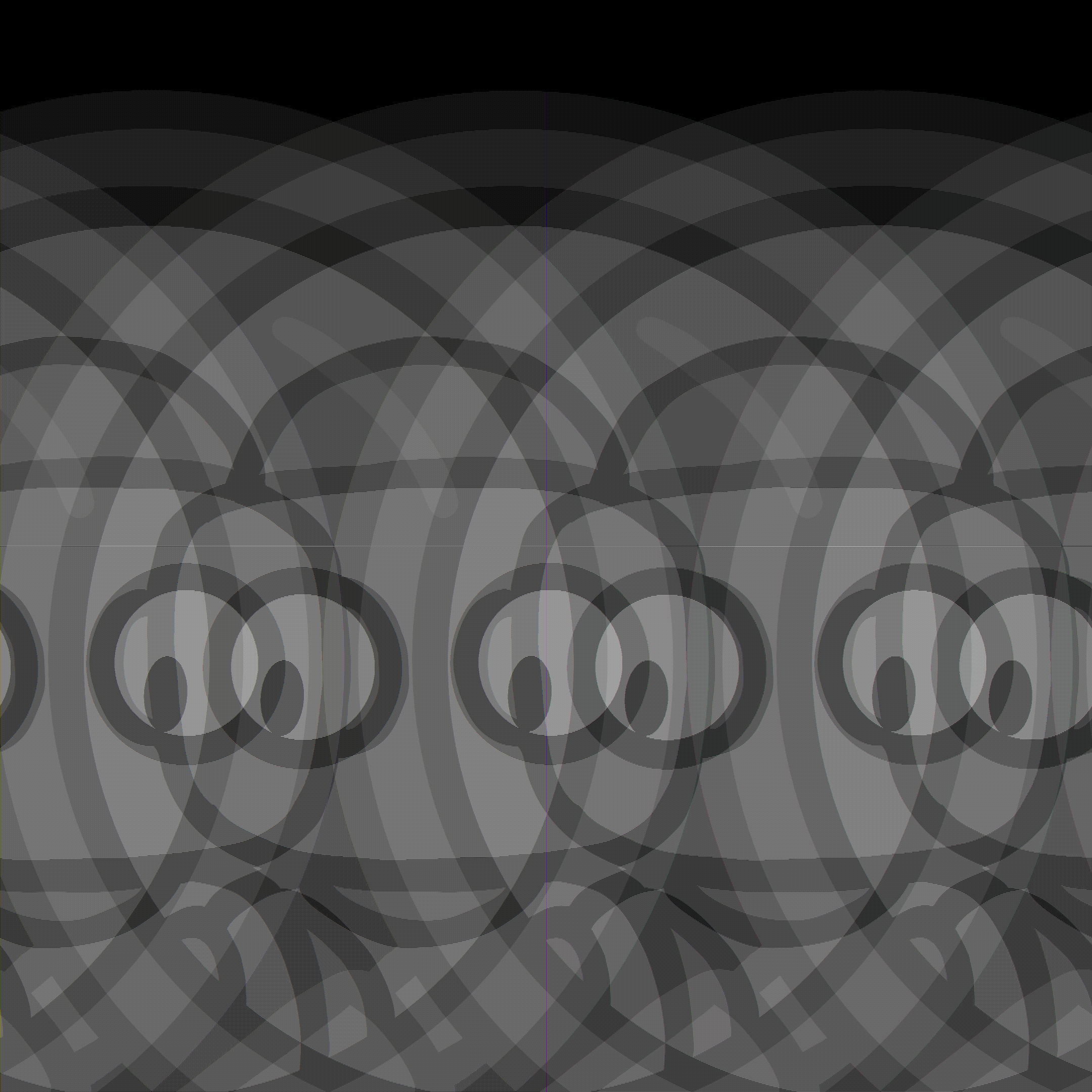}\\[-0.3cm]
\textbf{\normalsize IdeaSearch Collaboration}}

\author[1]{Zhuo-Yang Song}
\author[1]{Zeyu Cai}
\author[1]{Shutao Zhang}
\author[1]{Jiashen Wei}
\author[1,2]{Jichen Pan}
\author[1,3]{\mbox{Shi Qiu}}
\author[1,2]{Qing-Hong Cao}
\author[4]{Tie-Jiun Hou}
\author[5,6]{Xiaohui Liu}
\author[7]{Ming-xing Luo}
\author[1,2]{\mbox{Hua Xing Zhu}}

\affil[1]{School of Physics, Peking University, Beijing 100871, China}
\affil[2]{Center for High Energy Physics, Peking University, Beijing 100871, China}
\affil[3]{University of California, Berkeley}
\affil[4]{School of Nuclear Science and Technology,
University of South China, Hengyang, Hunan 421001, China}
\affil[5]{Center of Advanced Quantum Studies, School of Physics and Astronomy, Beijing Normal University, Beijing, 100875, China}
\affil[6]{Key Laboratory of Multi-scale Spin Physics, Ministry of Education, Beijing Normal University, Beijing 100875, China}
\affil[7]{Beijing Computational Science Research Center, Beijing 100193, China}

\begin{document}

\maketitle

\begin{abstract}
Symbolic regression (SR), the automated discovery of mathematical expressions from data, is a cornerstone of scientific inquiry. However, it is often hindered by the combinatorial explosion of the search space and a tendency to overfit. Popular methods, rooted in genetic programming, explore this space syntactically, often yielding overly complex, uninterpretable models.
This paper introduces IdeaSearchFitter, a framework that employs Large Language Models (LLMs) as semantic operators within an evolutionary search. By generating candidate expressions guided by natural-language rationales, our method biases discovery towards models that are not only accurate but also conceptually coherent and interpretable.
We demonstrate IdeaSearchFitter's efficacy across diverse challenges: it achieves competitive, noise-robust performance on the Feynman Symbolic Regression Database (FSReD), outperforming several strong baselines; discovers mechanistically aligned models with good accuracy-complexity trade-offs on real-world data; and derives compact, physically-motivated parametrizations for Parton Distribution Functions in a frontier high-energy physics application. IdeaSearchFitter is a specialized module within our broader iterated agent framework, IdeaSearch, which is publicly available at \href{https://www.ideasearch.cn/}{https://www.ideasearch.cn/}.
\end{abstract}

\section{Introduction}
\label{sec:introduction}

Symbolic regression~(SR), the task of inferring mathematical expressions from observational data, lies at the heart of scientific discovery. From Kepler's laws of planetary motion to Balmer's formula for spectral lines of hydrogen, the ability to distill concise, interpretable models from noisy measurements has driven paradigm-shifting insights across disciplines. In the scientific domain, SR has been explored as an interpretable machine learning model for data-driven discovery~\cite{gerwin,langley,falkenhainer,bacon1,10.5555/29379,doi:10.1126/science.1165893}. 
Yet, despite its foundational role, SR grapples with computational and statistical challenges: navigating an exponentially vast space of candidate expressions while avoiding overfitting to data artifacts rather than underlying mechanisms, see e.g.~\cite{srbench,makke2024interpretable} for recent reviews and the related benchmark~\cite{Orzechowski2018WhereAW,srbench,Aldeia2022InterpretabilityIS}.

A very successful approach to SR is the genetic programming paradigm~\cite{10.5555/1623755.1623877,10.5555/892491,koza}, which performs an evolutionary search over syntactic expression trees via random mutations and crossovers of predefined operators. This has led to many successful applications~\cite{koza,keijer,vladislavleva,Korns2011,Uy2010SemanticallybasedCI,jin,dsr,ffx,gpasr,ITEA,mrgp,lacavagp,livingreview}, as well as commercial tools like Eureqa~\cite{Dubcakova:2011:GPEM}.
While elegant in principle, these methods can face challenges in practice due to two intertwined limitations. First, the combinatorial explosion of possible expressions renders exhaustive exploration infeasible, even with modern hardware. 
Second, unconstrained syntactic variation can generate overly complex forms that capture spurious correlations, yielding models with low training error but poor generalization and interpretability. This overfitting problem creates a basic trade-off: more accurate models usually need to be more complex, but that makes them harder to understand and less likely to reflect real scientific mechanisms, which are important for validation.

Recent advances have partially alleviated these issues through multi-objective optimization and domain-informed priors, and the 
integration of neural networks to guide search~\cite{aifeynman,DBLP:journals/corr/MartiusL16,dsr,dsrgp,metamodel,transformers,Biggio2021NeuralSR,sindyae}.
One of the most successful examples is PySR~\cite{pysr}, which exposes a Pareto frontier of accuracy-complexity trade-offs, enabling users to select interpretable models from a spectrum of candidates. Physics-aware methods, such as AI-Feynman~\cite{aifeynman,Udrescu2020AIF2}, incorporate dimensional analysis and heuristic templates to bias toward compact, unit-consistent forms, succeeding on benchmark problems with known ground-truth equations. However, these approaches remain constrained: purely syntactic searches (e.g., Operon~\cite{operon}) risk irrelevance in domain-specific contexts, while rigid priors may overlook emergent complexities in real-world data. Collectively, they underscore a critical gap: not merely in \emph{how} to search, but in \emph{where} to search: the hypothesis space itself must be semantically bounded to favor expressions that are not only mathematically valid but also conceptually coherent with domain knowledge.

This work presents IdeaSearchFitter, a new SR framework inspired by FunSearch and AlphaEvolve~\cite{funsearch,alphaevolve}, that aims to shift the focus of the search from syntax toward semantics.  At its core lies the premise that effective discovery of interpretable models demands exploration within an interpretable ansatz space, a curated collection of candidate hypotheses motivated by explicit, natural-language rationales grounded in physical priors (e.g., conservation laws, boundary conditions, symmetry principles). LLMs serve dual roles as ansatz generators and reasoning agents: they articulate concise, testable justifications before proposing symbolic sketches, which are then canonicalized and refined. This "explain-then-formalize" workflow, embedded in a multi-island evolutionary loop, helps ensure that variation operators produce semantically meaningful mutations, reducing the search space while advancing the Pareto frontier in a higher-dimensional objective landscape that jointly optimizes accuracy, complexity, and interpretability.

Empirically, we demonstrate competitive performance on the Feynman Symbolic Regression Database (FSReD), outperforming several strong baselines~\cite{srbench}, with recovery rates exceeding 80\% under noise-free conditions and graceful degradation to 71.7\% at high noise levels ($\gamma=0.1$), outperforming baselines like PySR and AI-Feynman in robustness and efficiency.

Moving beyond the standard benchmarks, we also demonstrate IdeaSearchFitter's ability to discover powerful ansatz directly from complex, real-world datasets. We apply IdeaSearchFitter to real-world datasets from the Penn Machine Learning Benchmarks (PMLB)~\cite{pmlb}, identifying mechanistically aligned models (e.g., logarithmic memory effects for agricultural yields; regime-separated dynamics for supernova light curves) that achieve good Normalized Mean Squared Error (NMSE)-complexity trade-offs without overfitting. In another frontier application to high-energy physics, the framework identifies a simple equation that explains the complex internal structure of a proton from extracted data. This new equation was remarkably effective in predicting how the proton would behave in high-energy situations, demonstrated by the good $\chi^2/\mathrm{ndf}$.

Our approach has been integrated into a user-friendly iterated AI agent framework named IdeaSearch (see in \href{https://www.ideasearch.cn/}{https://www.ideasearch.cn/}), which allows diverse physical applications of agentic AI beyond SR~\cite{Song:2025pwy,Cao:2025shc}. We refer to IdeaSearch as an iterated agent framework where LLM-driven proposal-and-refinement operators produce formal hypothesis trajectories that are iteratively re‑fed for further improvement (see Section~\ref{sec:methodology} for a concise formalization).

The remainder of this paper is organized as follows. Section~\ref{sec:methodology} details the architecture of IdeaSearchFitter, from preprocessing and LLM-centric evolution to Pareto postprocessing. Section~\ref{sec:benchmarks} presents benchmark results on FSReD, including LLM phenotype characterization. Section~\ref{sec:applications} explores real-world PMLB datasets and the PDF case study. We conclude in Section~\ref{sec:conclusion} with implications and future directions.

\section{IdeaSearchFitter: A Natural Language Guided Search over Interpretable Ansatz}
\label{sec:methodology}

The primary challenge in SR is navigating a virtually infinite space of mathematical expressions, a task where popular genetic programming approaches face a combinatorial explosion and a tendency toward overfitting. While existing tools manage an accuracy-complexity tradeoff, they highlight a deeper question: the key is not just how we search, but where we search. To make this distinction precise, we formalize discovery as an iterative search driven by an operator, $\mathcal{T}$, that generates trajectories of hypotheses \footnote{a rigorous formulation is provided in Appendix~\ref{app:agent_formalism}}. Within this framework, we can understand traditional methods as employing a syntactic operator, $\mathcal{T}_{\text{syntactic}}$, which explores a vast syntactic space of all structurally valid expressions. In contrast, we argue that a more effective process would use a semantic operator, $\mathcal{T}_{\text{semantic}}$, to navigate a constrained semantic space of scientifically coherent hypotheses. While designing such an operator by hand is exceptionally difficult, Large Language Models offer a practical bridge, serving as a powerful approximation by generating hypotheses guided by natural-language rationales.

\begin{figure}
    \centering
    \includegraphics[width=\linewidth]{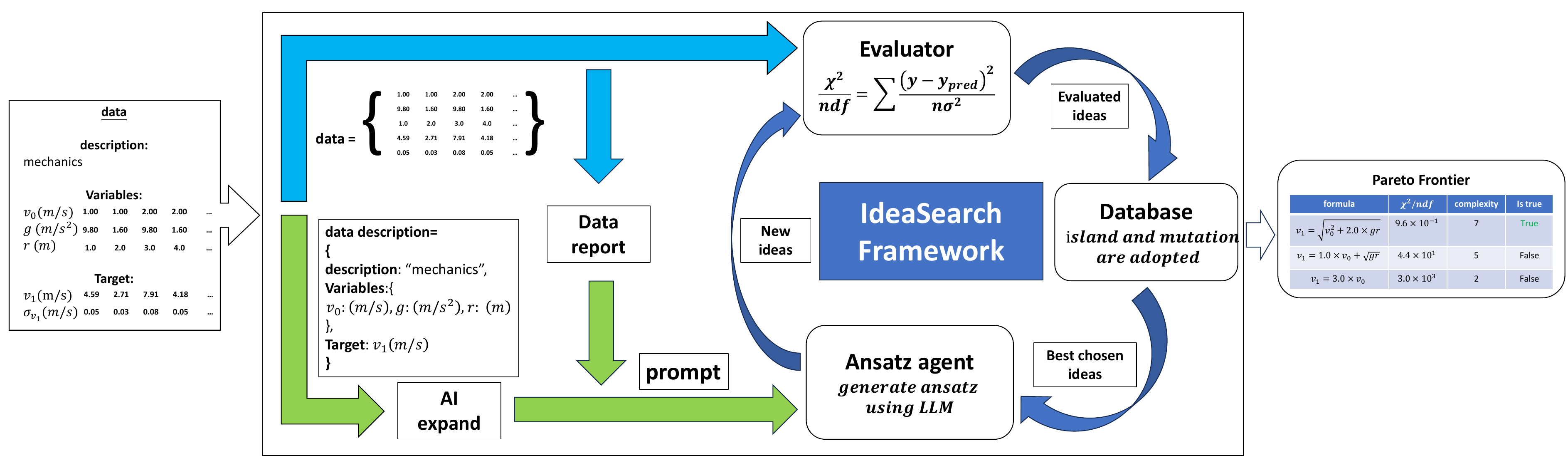}
    \caption{\textbf{Schematic overview of the IdeaSearchFitter framework.} The workflow begins with input data comprising variables ($g(m/s^2)$, $r(m)$), targets ($\nu(m/s)$), and uncertainties ($\Delta\nu(m/s)$), alongside a natural language description of the underlying mechanics (e.g., gravitational effects). This is condensed into a structured data description, an LLM will expand it to be an enriched context. An ansatz agent then prompts an LLM to generate candidate symbolic expressions, which are evaluated using the reduced $\chi^2/\mathrm{ndf}$ statistic ($\chi^2/\mathrm{ndf} = \sum [(y_\mathrm{pred,i} - y_\mathrm{true,i})^2/\sigma_i^2] / \mathrm{ndf}$) against the data. Evaluated ideas populate a multi-island database, where local mutations and inter-island migrations promote diversity and prevent premature convergence. Promising ansatz are iteratively refined via LLM-guided polishing and global optimization. The search culminates in Pareto frontier selection, trading off fit quality ($\chi^2/\mathrm{ndf}$) against structural complexity (e.g., expression tree nodes), yielding a set of interpretable models; the true relation (e.g., $\nu = \sqrt{2gr}$) is highlighted among candidates.}
    \label{fig:flow chart}
\end{figure}

Our work, IdeaSearchFitter, operationalizes this insight. It employs LLMs as semantic mutation and crossover operators within a multi-island evolutionary loop, shifting the search from syntactic trees to an interpretable ansatz space. Each candidate is accompanied by an explicit, human-readable rationale, ensuring that variation is conceptually meaningful. While our framework is compatible with hybrid approaches, the experiments in this paper exclusively use this LLM-driven loop to demonstrate its distinct effect. This shift from syntactic to semantic search is central to IdeaSearchFitter's advantage: by privileging conceptually simple, domain-motivated hypotheses, the tool reduces overfitting and advances the Pareto frontier in a space that jointly considers accuracy, complexity, and interpretability. We now turn from this abstract rationale to the concrete implementation (also shown in Fig.~\ref{fig:flow chart}). For a practical guide and tutorials, readers are referred to \href{https://www.ideasearch.cn/docs/fitter/}{https://www.ideasearch.cn/docs/fitter/}.

\subsection{Preprocessing}
The preprocessing stage fuses two complementary information streams into a dynamic prompt for the evolutionary loop. First, it processes structured metadata and user-provided domain priors (e.g., boundary conditions), which an LLM expands into an enriched, internally consistent contextual summary. Concurrently, it ingests numerical data and their uncertainties to compute key diagnostics and lightweight reports. These two streams are then fused into a single contextual prompt that is supplied to the generation agents. Critically, at each epoch, this prompt is augmented with a snapshot of the current top-performing ideas, providing high-quality, evolving seeds for the LLMs.

\subsection{Evolutionary search loop}
The search is implemented as a multi-island evolutionary process whose variation operators are intentionally LLM-centric. In the reported experiments, all new symbolic candidates are proposed by LLM agents; classical tree mutation and crossover operators remain available in the codebase for future hybrid configurations but are disabled here. Diversity and variation are thus induced by controlled LLM generation, by leveraging heterogeneous LLM backends, by seeding from distinct prior ideas, and by conditioning on island-specific contexts.

Each candidate is produced by a two-stage ansatz agent~\cite{stepbystep}. A proposal generator LLM first returns a short natural-language rationale together with one or more sketch expressions. An extractor LLM then converts the sketch into a single canonical, parseable symbolic expression with associated metadata such as suggested parameter names, initial bounds, and implied units. This decomposition enables broad semantic exploration while enforcing syntactic validity and canonicalization.

Each island evolves its ideas independently.  Periodically, islands are repopulated with the global best ideas. Tree mutation and crossover are deactivated in the experiments, so structural variation arises from fresh LLM proposals conditioned on local populations and global best ideas.

\subsection{Postprocessing}
After the search completes (or upon user interruption), the archive is analyzed to produce a compact, actionable report. The primary postprocessing step is Pareto frontier construction over two axes: fit quality (reduced $\chi^2/\mathrm{ndf}$) and structural complexity (canonical node count of the expression tree).~\ref{subsec:protocol}. The reported Pareto set is presented together with the natural‑language rationales attached to each surviving idea, parameter estimates with uncertainties, unit checks, and diagnostic plots (training/validation residuals, simple extrapolation checks). This Pareto‑driven report gives human experts a concise set of interpretable candidates that trade off accuracy and complexity; further manual refinement is left to domain specialists as needed.

Implementation details and additional algorithmic options (e.g., alternative distributor schedules, explicit mutation/crossover modes, full prompt texts, and exact hyperparameters) are collected in Appendix~\ref{app:ideasearch_impl}.

\section{Benchmark Performance and a Dynamic Methodology for LLM Evaluation}
\label{sec:benchmarks}

After introducing IdeaSearchFitter's approach of searching within an interpretable ansatz space (Section~\ref{sec:methodology}), we now move on to systematic empirical validation. To show that our theoretical framework leads to measurable performance improvements, we tested IdeaSearchFitter on the FSReD~\cite{srbench}, which is a widely used dataset in SR. FSReD includes 120 problems with known ground-truth expressions, allowing for fair and reproducible comparisons with other top methods.

Beyond establishing competitiveness on this reference benchmark, we leverage the unique architecture of IdeaSearchFitter to address a broader challenge in LLM evaluation. Standard benchmarks typically measure model performance through static final-answer accuracy metrics, which obscure the diversity of problem-solving strategies inherent in different models. We demonstrate that IdeaSearchFitter can serve as a dynamic evaluation framework for characterizing the distinct reasoning phenotypes of LLMs within a complex, iterative search process.

\subsection{Evaluation Protocol and Experimental Setup}
\label{subsec:protocol}

To ensure fair comparison, we adopt a standardized evaluation protocol. First, all baseline methods use the operator sets and hyper-parameters listed in Tab.~\ref{table:hyperparams2}, which were individually optimized for each competitor. Second, every method is allocated approximately the same computational budget, measured in wall-clock time. Finally, IdeaSearchFitter receives only the minimal natural-language description: variable names, target name, and units. For this section, IdeaSearchFitter is always run in \textit{fast} mode (see Tab.~\ref{table:hyperparams2}); the median words consumption reported in Tab.~\ref{tab:noise_results} confirms that this choice is reasonable.

Our assessment focuses on the following metrics:

\begin{itemize}
    \item \textbf{Recovery Rate:} The fraction of problems for which an expression algebraically equivalent to the ground truth is successfully recovered. Following the SRBench methodology~\cite{srbench}, equivalence is verified using SymPy: an expression $\hat{\phi}$ recovers the ground truth $\phi^*$ if their difference or ratio simplifies to a constant.

    \item \textbf{Noise parameter $\gamma$ (and SNR relation):} A measure of data quality, defined as the ratio of the signal variance to the noise variance, $\mathrm{SNR} = \sqrt{\frac{\sum_{i=1}^{N} (y_i - \bar{y})^2}{N \sigma^2}}$, where $\bar{y}$ is the mean of the target values and $\sigma$ is the noise standard deviation. In our experiments, additive Gaussian noise is introduced with $\sigma = \gamma \sqrt{\frac{\sum_{i=1}^{N} (y_i - \bar{y})^2}{N}}$, such that $\gamma = 1/\mathrm{SNR}$ represents the noise-to-signal ratio, enabling controlled assessment of robustness across varying data quality levels. we report results in terms of $\gamma$ (noise-to-signal ratio).

    \item \textbf{Median Epochs or 3/4 Quartile Epochs:} The distribution of epochs required to solve problems. A tight distribution indicates consistent efficiency, while a wide gap suggests the model can solve both easy and hard problems.
        
    \item \textbf{Median Words:} The typical length of reasoning generated in total per problem, reflecting the model's verbosity and reasoning style.
    
\end{itemize}

The following evaluation focused on these metrics as variables of noise intensity or different methods.

\subsection{Performance on the FSReD Benchmark}

We systematically evaluated IdeaSearchFitter on the FSReD dataset.  We conducted experiments with and without dimensional analysis (unit consistency checking) to assess performance under different levels of physical constraint enforcement. For comparison, all baselines were run with dimensional analysis turned off. Unless otherwise stated, the method refers to the version without dimensional analysis.

To ensure comparability with prior work, we introduced additive Gaussian noise following the SRBench methodology. The noise standard deviation is defined as:
\begin{equation}
\sigma = \gamma \sqrt{\frac{\sum_{i=1}^{N} (y_i - \bar{y})^2}{N}},
\end{equation}

where $\gamma$ is the noise-to-signal ratio and $\bar{y}$ is the mean of the target variable.

We determined the recovery rate algorithmically following Definition 4.1 from SRBench~\cite{srbench}. An expression $\hat{\phi}$ successfully recovers the ground truth $\phi^*$ if it is not a constant and satisfies one of two conditions: their difference simplifies to a constant ($\phi^* - \hat{\phi} = a$), or their ratio simplifies to a non-zero constant ($\phi^* / \hat{\phi} = b \neq 0$). Our implementation uses SymPy for symbolic parsing and simplification.

\begin{figure}[ht!]
\centering
\includegraphics[width=.55\linewidth]{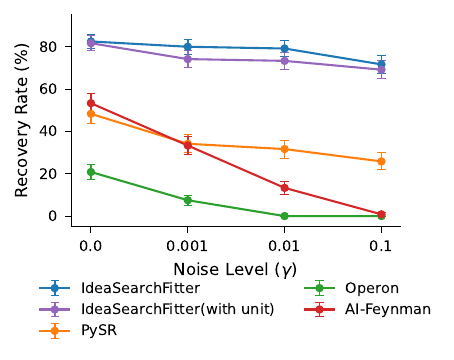}
\caption{\textbf{Recovery rate versus noise level on the FSReD.} The plot illustrates the fraction of ground-truth expressions successfully recovered by each method as a function of additive Gaussian noise intensity ($\gamma$). IdeaSearchFitter (blue line with error bars) demonstrates good robustness and ability, with recovery rates exceeding 70\% even at $\gamma=0.1$, compared to baselines including PySR (orange), Operon (green), and AI-Feynman (red). The purple line represents IdeaSearchFitter evaluated with dimensional analysis enabled. Error bars denote standard error across 120 problems.}
\label{fig:noise-robustness}
\end{figure}

In the standard setting without unit checking, IdeaSearchFitter demonstrates competitive performance with strong resilience to noise across all tested levels. As shown in Tab.~\ref{tab:fsred} and Fig.~\ref{fig:noise-robustness}, its recovery rate degrades only gradually with increasing noise intensity, consistently surpassing strong baselines such as PySR, Operon, and AI-Feynman.

\begin{table}[ht!]
    \centering
    \caption{\textbf{Recovery rates on the FSReD. PySR, AI-Feynman, and Operon is evaluated without dimensional analysis.}}
    \label{tab:fsred}
    \begin{tabular}{lcccc}
    \toprule
         \textbf{Noise Level ($\gamma$)} &  0.0 &  0.001 &  0.01 & 0.1\\
    \midrule
         \textbf{IdeaSearchFitter} &  \textbf{82.5\%} &  \textbf{80.0\%} &  \textbf{79.2\%} & \textbf{71.7\%}\\
 IdeaSearchFitter(with unit)& 81.7\%& 74.2\%& 73.3\%&69.2\%\\
         PySR             &  48.3\% &  34.2\%&  31.7\%   & 25.8\% \\
         Operon           &  20.8\% &  7.5\% &   0.0\% & 0.0\%\\
         AI-Feynman       &  53.3\% &  33.3\% &  13.3\% & 0.8\%\\
    \bottomrule
    \end{tabular}
\end{table}

To quantify search efficiency, we note that IdeaSearchFitter routinely attains recovery probabilities above 0.7 within 10 minutes of wall-clock time. This indicates a rapid convergence to high-quality solution spaces. The efficiency stems from the parallelized LLM-guided proposals and server-side API acceleration, which usually enable IdeaSearchFitter to find satisfactory solutions without needing to complete all search epochs.

When dimensional consistency (unit checking) is enforced, IdeaSearchFitter retains comparable performance, particularly in noisy regimes. The semantic guidance inherent in natural-language rationales largely subsumes the benefits of explicit unit checks, yielding only modest differences in recovery rates (Tab.~\ref{tab:fsred})~\footnote{The recovery rates for AI-Feynman are extracted from SRBench~\cite{srbench}.}. However, enabling unit consistency can further accelerate convergence for select problems, reducing the 3/4 quartile epochs to solution (e.g., from 10.00 to 3.50 epochs under $\gamma=0.1$ noise; cf. Tab.~\ref{tab:noise_results}). This suggests that while LLM-driven priors effectively incorporate dimensional heuristics, supplementary checks can provide a lightweight refinement for targeted efficiency gains.

\begin{table}[ht!]
\centering
\caption{\textbf{Performance metrics for IdeaSearchFitter under varying noise levels, with and without dimensional analysis. Metrics include median words used and 3/4 quartile epochs to the solution.}}
\label{tab:noise_results}
\begin{tabular}{lccc}
\toprule
\textbf{Method}& \textbf{Noise Level($\gamma$)}& \textbf{Median Words} & \textbf{3/4 Quartile} \\
\midrule
\multirow{4}{*}{IdeaSearchFitter} 
& 0.0& 1648.0& 2.00  \\
& 0.001& 2822.5& 7.00  \\
& 0.01& 2235.0& 8.00  \\
& 0.1& 3757.0& 10.00 \\
\midrule
\multirow{4}{*}{IdeaSearchFitter(with unit)} 
& 0.0& 1538.0& \textbf{2.00} \\
& 0.001& 2233.0& \textbf{4.00} \\
& 0.01& 2941.0& \textbf{4.25} \\
& 0.1& 1925.0& \textbf{3.50} \\
\bottomrule
\end{tabular}
\end{table}

In summary, IdeaSearchFitter proves to be a robust and efficient SR tool in our evaluation. Its results on the FSReD benchmark empirically affirm the value of LLM-driven, semantically constrained search as an effective paradigm for scientific model discovery. The method's notable tolerance to noise is especially pertinent for practical applications, where observational uncertainties are commonplace. A detailed benchmark characterizing the performance of different LLMs as the generative engine within our framework is presented in Appendix~\ref{app:llm_benchmark}.

The subsequent analysis focuses on NMSE-complexity trade-offs and includes representative case studies that highlight the generation of mechanistically aligned structures. This demonstrates how the principles of our framework translate to discovering compact, interpretable models that are essential for scientific insight.

\section{Applications to Real-World and Frontier Scientific Problems}
\label{sec:applications}

While performance on benchmarks with known ground-truth solutions demonstrates algorithmic soundness, the ultimate measure of a symbolic discovery tool is its ability to extract novel, meaningful, and interpretable models from noisy, real-world observational data. This section transitions from the controlled environment of benchmarks to the frontier of scientific applications. We first assess IdeaSearchFitter's performance on a curated set of real-world problems from the PMLB repository~\cite{pmlb} and then apply it to two challenging, open problems in high-energy physics.

\subsection*{Evaluation Protocol on Real-World Datasets}

For datasets without a known ground-truth formula, our evaluation protocol shifts to assessing the trade-off between predictive accuracy and model simplicity—a cornerstone of scientific modeling. Each dataset is split into a 75\% training set and a 25\% validation set. All methods generate candidate expressions on the training data, and the final models are evaluated on the validation data. Our analysis is centered on the Pareto frontier of solutions in the accuracy-complexity plane, using the following metrics:

\begin{itemize}
    \item \textbf{Normalized Mean Squared Error (NMSE):} To measure predictive accuracy on given data, we use the NMSE, defined as:
    $$
    \text{NMSE} = \frac{\sum_{i}(y_i - y_{\text{pred},i})^2}{\sum_{i}(y_i - \bar{y})^2}
    $$
    where \(\bar{y}\) is the mean of the true target values in the test set. An NMSE less than 1.0 indicates performance better than simply predicting the mean.

    \item \textbf{Expression Complexity:} Used as a proxy for interpretability, complexity is measured by the number of nodes in the expression's syntax tree.
\end{itemize}

\subsection{Discovering Interpretable Models from Observational Data in PMLB}
\label{ssec:pmlb}

We curated 8 real-world regression tasks from the PMLB repository~\cite{pmlb} via a rigorous, reproducible three-stage filtering pipeline (detailed in Appendix~\ref{app:dataset_selection}). These datasets, spanning biology, astronomy, and fluid dynamics, lack known ground-truth equations and emphasize observational phenomena amenable to symbolic discovery. The process yields datasets suitable for uncovering concise, universal symbolic laws under realistic noise and variability. On these eight tasks, IdeaSearchFitter is run in \textit{expert} mode (Tab.~\ref{table:hyperparams2}); dataset descriptions are taken from the PMLB metadata. As detailed in Appendix~\ref{app:dataset_selection}, we manually verified that these descriptions contain no explicit or implicit hints of the underlying ground-truth expressions.

We adopt a Pareto-based methodology in the (NMSE, complexity) plane. Each dataset is randomly split into 75\% training and 25\% testing; methods independently generate candidate expressions on the training split. Predictive accuracy is quantified by the Normalized Mean Squared Error (NMSE) on the validation split, and model complexity by the number of nodes in the expression tree. We compare only solutions lying on the Pareto frontier, which constrains overfitting while favoring concise, interpretable forms and is consistent with the selection and refinement strategy outlined in Section~\ref{sec:methodology}.

Generally, in our tests, IdeaSearchFitter achieves comparable NMSE to PySR across most datasets with reduced or similar expression complexity, maintaining a balanced accuracy–complexity trade-off. Although Operon occasionally attains marginally lower NMSE, it often does so at higher complexity, compromising interpretability, while AI-Feynman struggles in small-sample or high-dimensional datasets.

After this subsection, detail NMSE-complexity profiles and case studies, demonstrating mechanism-aligned structures via IdeaSearchFitter, free of overfitting artifacts.

\subsubsection{NMSE performance}

\begin{figure}
    \centering
    \includegraphics[width=.8\linewidth]{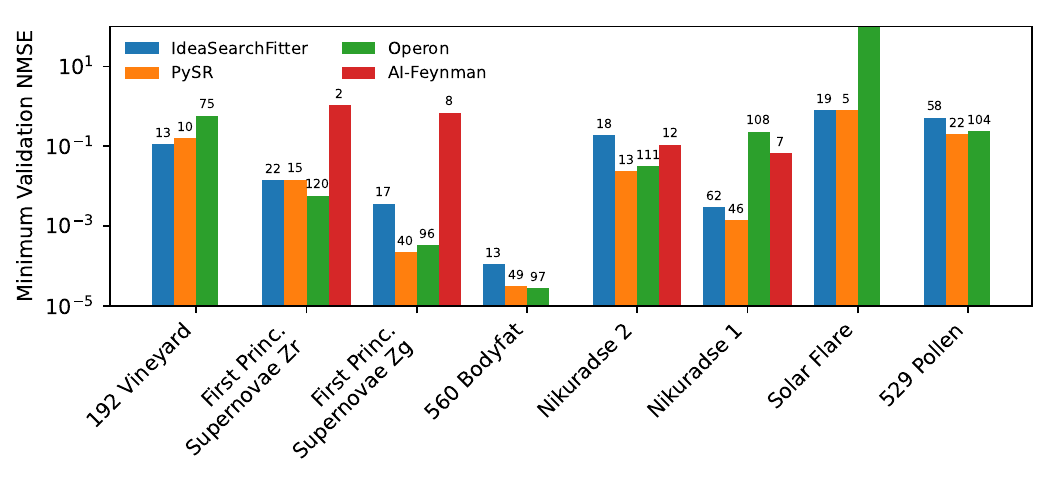}
    \caption{\textbf{Minimum validation NMSE achieved by SR methods across eight PMLB datasets.} Bars: lowest validation NMSE and complexity for IdeaSearchFitter (blue), PySR (orange), Operon (green), AI-Feynman (red); absent bars indicate failures (e.g., AI-Feynman on 192\_vinyard datasets).}
    \label{fig:NMSE}
\end{figure}

Fig.~\ref{fig:NMSE} displays the minimum validation NMSE and corresponding expression complexity achieved by each method across eight representative real-world PMLB datasets. Missing bars for AI-Feynman indicate its failure to produce a valid expression, usually due to incompatibilities with small sample sizes or high dimensionality in its neural network preprocessing.

We select the minimum validation NMSE as a global performance indicator to showcase fitting superiority, as it captures each method's best achievable generalization on held-out data, independent of specific complexity levels. This metric facilitates direct cross-method comparisons and underscores the balance between accuracy and overfitting risk, aligning with our framework's multi-objective ethos (Section~\ref{sec:methodology}).
Here, we adopt the minimum validation NMSE, i.e., the lowest NMSE observed on the held-out validation set across the entire complexity trajectory, as the primary figure of merit. This scalar is not exposed to the optimiser during symbolic fitting; consequently, its value can increase under either under- or over-fitting conditions. By recording the validation NMSE of every terminal model on the Pareto frontier, we obtain an unbiased, post-hoc measure of how effectively a method expands the Pareto frontier while controlling over-fitting, thereby satisfying the multi-objective criterion outlined in Section~\ref{sec:methodology}.

As shown in Fig.~\ref{fig:NMSE}, IdeaSearchFitter achieves comparable validation NMSE with PySR on 6 of 8 datasets, comparable to PySR overall: on 3 datasets, its NMSE is lower than PySR's; on the remaining 3, NMSE is higher but with substantially lower complexity on 2. The method fails to produce comparably effective expressions on the other 2 datasets, which is consistent with challenges in highly variable observational data. Relative to Operon and AI-Feynman, IdeaSearchFitter occupies a good position on the accuracy-complexity Pareto frontier, obviously, as Operon secures marginally lower NMSE only at markedly higher complexity, and AI-Feynman underperforms on both NMSE and overfitting mitigation. In contrast, IdeaSearchFitter's Pareto-screened candidates maintain consistent trade-offs without overfitting artifacts, as detailed in the following case studies.

Validation NMSE profiles of IdeaSearchFitter exhibit smooth monotonic decrease without rebounds (Fig.~\ref{fig:example}), aligning with the noise robustness observed on FSReD (Fig.~\ref{fig:noise-robustness}). Isolated failures (e.g., on nikuradse\_2 and 529\_pollen) fall within expected ranges for noisy observational data. The following subsection provides detailed case studies, including complexity-NMSE curves and structural analyses, to further elucidate stability under constrained complexity and the emergence of interpretable, mechanism-aligned expressions in IdeaSearchFitter.

\subsubsection{Case Studies: Interpretability and Non-Overfitting Behavior}

To illustrate a key advantage of IdeaSearchFitter: generating stable, mechanistically interpretable expressions under comparable complexity and interoperability budgets. We examine two representative PMLB datasets: vineyard yield and supernova light curves. Rationales for the highlighted models derive from the system's "explain-then-formalize" workflow (Section~\ref{sec:methodology}), where each iteration yields a natural-language justification prior to symbolic instantiation. This process ensures that structures and parameters align with domain priors, facilitating scientific insight.

\begin{figure}[ht]
\centering
\begin{subfigure}[b]{0.45\linewidth}
  \includegraphics[width=\linewidth]{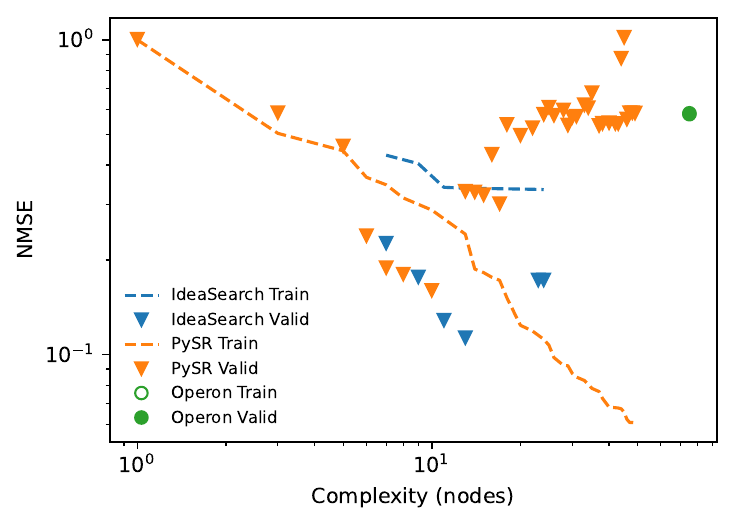}
  \caption{Vineyard yield (192\_vineyard): complexity versus NMSE.}
  \label{fig:192_vineyard}
\end{subfigure}\hfill
\begin{subfigure}[b]{0.45\linewidth}
  \includegraphics[width=\linewidth]{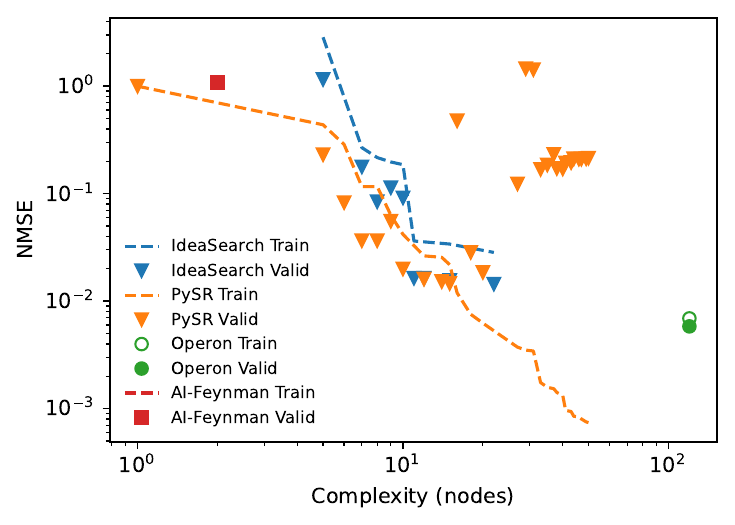}
  \caption{Supernova light curve (first\_principles \_supernovae\_zr): complexity versus NMSE.}
  \label{fig:first_principles_supernovae_zr}
\end{subfigure}
\caption{\textbf{Complexity-NMSE Pareto frontiers for two PMLB datasets (75\%-25\% train-validation split).} Training NMSE (dashed) and validation NMSE (markers) versus tree nodes. IdeaSearchFitter (blue) shows monotonic decrease without overfitting rebounds; maps to mechanisms (logarithmic memory in (a); $|t|$-regime separation in (b); cf. Tab.~\ref{tab:publication_comparison}). Baselines: PySR (orange, rebound at $\sim10$ nodes in (a)); Operon (green); AI-Feynman (red, failures).}
\label{fig:example}
\end{figure}

\textbf{Vineyard yield: Domain intuition and structural mapping.} Agricultural systems often exhibit interannual memory and saturation effects, suggesting a form incorporating logarithmic compression for diminishing returns, weighted prior-year contributions, and a baseline offset. At moderate complexity, IdeaSearchFitter recovers the concise model
\begin{equation}
\hat{y} = \alpha \log\left( \frac{9}{2} \text{lugs}{1989} + \text{lugs}{1990} \right) - \beta,
\end{equation}
where the logarithm encodes saturation, the $9/2$ weight rescales historical memory (aligning with unconstrained fits yielding $c \approx 4.6$, interpreted as a normalization factor), and $\beta$ sets the baseline. This form mirrors agronomic memory-recovery mechanisms and, at equivalent complexity, shows decreasing training NMSE with stabilizing validation NMSE (Fig.~\ref{fig:example}; cf. Tab.~\ref{tab:publication_comparison}).

\textbf{Supernova light curve: Physical priors and structural mapping.} Asymmetric "fast rise, slow decay" dynamics are captured by
\begin{equation}
\hat{y} = \exp\left( -\alpha \left( \frac{t - |t|}{2} \right)^2 - \beta \left( \frac{t + |t|}{2} \right) \right),
\end{equation}
with $\frac{t - |t|}{2} = \min(t, 0)$ modeling a near-Gaussian pre-peak rise (width $\alpha \approx 8.8 \times 10^{-3}$) and $\frac{t + |t|}{2} = \max(t, 0)$ an exponential post-peak decay (rate $\beta \approx 4.3 \times 10^{-2}$). The parsimonious parameterization separates regimes explicitly, assigning physical meaning to coefficients and achieving low validation NMSE at constrained complexity (Tab.~\ref{tab:publication_comparison}; Fig.~\ref{fig:example}).

\begin{table}[htbp]
\centering
\caption{\textbf{Best-performing models on two PMLB datasets.} Fitted equations, complexity, train/validation NMSE (Pareto minimum validation NMSE).}
\label{tab:publication_comparison}
\begin{tabular}{lp{0.42\textwidth}rrr}
\toprule
\textbf{Method} & \textbf{Fitted Equation} & \textbf{Cplx.} & \textbf{Train NMSE} & \textbf{Val. NMSE} \\
\midrule
\multicolumn{5}{l}{\textbf{Vineyard}(Target:$\text{lugs}_{1991}$)} \\
IdeaSearch & $6.93 \cdot \log\left(\frac{9}{2}\text{lugs}_{1989} + \text{lugs}_{1990}\right) - 3.27$ & 13 & 0.338 & 0.113 \\
PySR & $(14.1 - \text{lugs}_{1990}) \times (\sqrt{\text{lugs}_{1989}} - 1.84) + 19.7$ & 10 & 0.288 & 0.160 \\
Operon & Unsimplified model & 75 & 0.580 & 0.585 \\
 AI-Feynman& Failed& NaN& NaN&NaN\\
\midrule
\multicolumn{5}{l}{\textbf{First Principles Supernova Zr}(Target:$I$)} \\
IdeaSearch & $
e^{-0.0215\cdot(t + |t|) - 2.2\times 10^{-3}\cdot(t - |t|)^2 }
$ & 22 & 0.0284 & 0.0144 \\
PySR & $
0.969/\sqrt{e^{0.0834\cdot(t + \cos (t))} + 3.8 \times 10^{-5}\cdot e^{-t}}
$ & 15 & 0.0218 & 0.0146 \\
Operon & Unsimplified model & 120 & 0.0069 & 0.0058 \\
AI-Feynman& $0.0469\pi$& 2& 1.14&1.09\\
\midrule
\end{tabular}
\end{table}

As summarized in Tab.~\ref{tab:publication_comparison} and Fig.~\ref{fig:example}, IdeaSearchFitter's models demonstrate good interpretability at matched complexity: the vineyard's weighted logarithmic memory and supernova's $|t|$-derived regime separation map directly to domain mechanisms, unlike baselines' nested, opaque operators. This non-overfitting NMSE-complexity profile arises from two core features (Section~\ref{sec:methodology}): (i) semantically guided proposals biasing toward mechanism-aligned primitives; and (ii) multi-island diversity sustaining distinct semantic lineages, averting dominance by high-complexity, low-interpretability fits. These case studies underscore the framework's efficacy in yielding actionable scientific models from observational data.

The consistent ability of IdeaSearchFitter to derive compact, mechanistically plausible models from noisy observational data motivates its application to pressing challenges in fundamental science, where such models can provide critical physical insights.

\subsection{Case Study in High-Energy Physics: Extracting parton distributions inside a proton}
\label{ssec:pdf}

High-energy physics presents a fertile ground for symbolic discovery. While the Standard Model provides a robust theoretical framework, many crucial experimentally measurable quantities lack first-principle analytical forms. This necessitates the use of phenomenological models and flexible parametrizations to bridge theory and experiment. SR, with its ability to distill complex data into compact, interpretable analytical forms, offers a powerful tool in this context: not just for discovering new relationships, but for finding more insightful and robust representations of established physical quantities~\cite{mlsymbolic,cranmer2020discoveringsymbolicmodelsdeep,Dong:2022trn,Thongkonsing:2023xxc,Butter_2024,Tsoi:2024pbn,Soybelman:2024mbv,AbdusSalam:2024obf,dotson2025generalizedpartondistributionssymbolic,Bendavid:2025urn,Vent:2025ddm,Bahl:2025jtk,Yuan:2025eyi,moralesalvarado2025foundationmodelsequationdiscovery}.

A prime example is understanding the inner structure of the proton.
According to quantum chromodynamics (QCD), the proton is not just a simple trio of two up quarks and one down quark. It is a roiling sea of transient quark-antiquark pairs and the gluons that bind them. To make predictions for experiments at colliders like the Large Hadron Collider, physicists need a precise map of this internal landscape. This map is known as a PDF, first proposed by Feynman back in the 1960s.

A PDF, denoted as $f_i(x,Q)$, essentially describes the probability of finding a specific type of particle, or a "parton", inside the proton carrying a certain fraction $x$ of the proton's momentum at an energy scale $Q$. These functions are indispensable, but here is a catch: for $60$ years, their exact mathematical form is not known. In principle, their numeric values can be computed from QCD first principles, but is extremely challenging in practice. While the only certain thing we know about PDF is its evolution obeys the famous Dokshitzer-Gribov-Lipatov-Altarelli-Parisi (DGLAP) equations~\cite{DGLAP1,DGLAP2,DGLAP3}, via perturbative calculable splitting functions $P_{ij}(z,\alpha_s(Q))$:
\begin{equation}
\frac{\partial f_i(x,Q)}{\partial \log Q^2} = \int_x^1 \frac{dz}{z} P_{ij}\!\left(\frac{x}{z}, \alpha_s(Q)\right) f_j(z,Q),
\end{equation}
with $\alpha_s(Q)$ the strong coupling. 

Conventionally, physicists relied on flexible but complex parametrizations, fitting them to vast amounts of experimental data. While effective, these models can be hard to interpret physically and will become unreliable when extrapolated to regimes where data are unavailable.

This is where SR offers a powerful alternative. Rather than assuming a functional form, we can discover one. This work uses our IdeaSearchFitter methodology to analyze data of the proton's structure. The goal is not merely to fit the data, but to distill the complex dynamics into a compact, insightful, and robust analytical formula. By doing so, we aim to find a more fundamental representation of this key physical quantity, bridging the gap between raw data and the underlying principles of QCD.

Here, we apply SR, via IdeaSearchFitter (Section~\ref{sec:methodology}), to discover compact, interpretable ansatz directly from data (CT18NNLO grids). Fits target six light flavours ($u_v,d_v,s=\bar{s},\bar{u},\bar{d},g$).
Evaluation uses reduced $\chi^2/\mathrm{ndf}$. This setup bridges symbolic discovery with QCD phenomenology by performing 2D fits in the $x$-$Q$ plane to probe for empirical forms that approximate DGLAP evolution, assessing their extrapolation capability beyond training scales.

Having established the physical context, we demonstrate IdeaSearchFitter's efficacy on data generated from established fits, highlighting its extrapolation capabilities compared to baselines like PySR~\cite{pysr}.
Our approach directly targets physically motivated forms that approximate DGLAP evolution in the $x$-$Q$ space, in contrast to traditional methods that parametrize at a single scale and evolve numerically.
By discovering compact, interpretable 2D expressions $f_i(x,Q)$ from data, the framework produces empirical forms that could capture some of the scaling behaviors without the full perturbative machinery. Such forms provide compact emulators that respect physical constraints like positivity while extrapolating more reliably than other baseline methods.

The input data comprises a 2D grid of observables for each flavour $i \in {u_v, d_v, s, g, \bar{u}, \bar{d}}$, sampled over $x \in [0.001, 0.4]$ and $Q \in [2, 5]\,\mathrm{GeV}$, with nodes spaced logarithmically along both axes with 20 points per axis (totally 400 points for each method to search). The corresponding PDFs for each species are obtained from the CT18NNLO set~\cite{Hou:2019qau} using the \texttt{LHAPDF}~\cite{Buckley:2014ana} library, and the corresponding uncertainties of PDF values are estimated using the Hessian method at the 68\% confidence level. This kinematic window balances coverage of valence-dominated large-$x$ regions with sea-like small-$x$ behavior, while spanning scales from nonperturbative onset to perturbative applicability, with evaluation via reduced $\chi^2/\mathrm{ndf}$.

IdeaSearchFitter is run in expert mode (Tab.~\ref{table:hyperparams2}) for 60 epochs, with prompts involving DGLAP evolution. The search explores a semantically constrained ansatz space, biasing toward forms like $x^a(Q) (1-x)^{b(Q)}$ or $\log(1/x)^{\lambda(Q)} \exp(-c(Q) x)$. This allows the discovery process to be guided by data and physical principles, rather than being confined to a fixed functional basis chosen by experts. For comparison, we benchmark against PySR under matched computational budgets, using the same grid and hyper-parameters as shown in Tab.~\ref{table:hyperparams2}.

To probe extrapolation reliability, we evolve the reference CT18NNLO PDFs to $Q=100\,\mathrm{GeV}$ (well beyond the training range) and compare against held-out validation grids at this scale, quantifying agreement via integrated $\chi^2$ over $x$.

\paragraph{Results: Compact Ansatz and Improved Extrapolation Performance}

To assess generalization beyond random train–validation splits and to directly probe overfitting mitigation, we evaluate a two-dimensional PDF case by extrapolating from the training scales $Q\in[2,5]\,\mathrm{GeV}$ to a high validation scale $Q=100\,\mathrm{GeV}$. We compare IdeaSearchFitter with PySR as the strongest and most broadly adopted baseline identified earlier. All fits are performed on the same CT18NNLO data grids and scored by reduced $\chi^2$; model complexity is measured by canonical node count.

\begin{figure}[ht!]
\centering
\includegraphics[width=.9\linewidth]{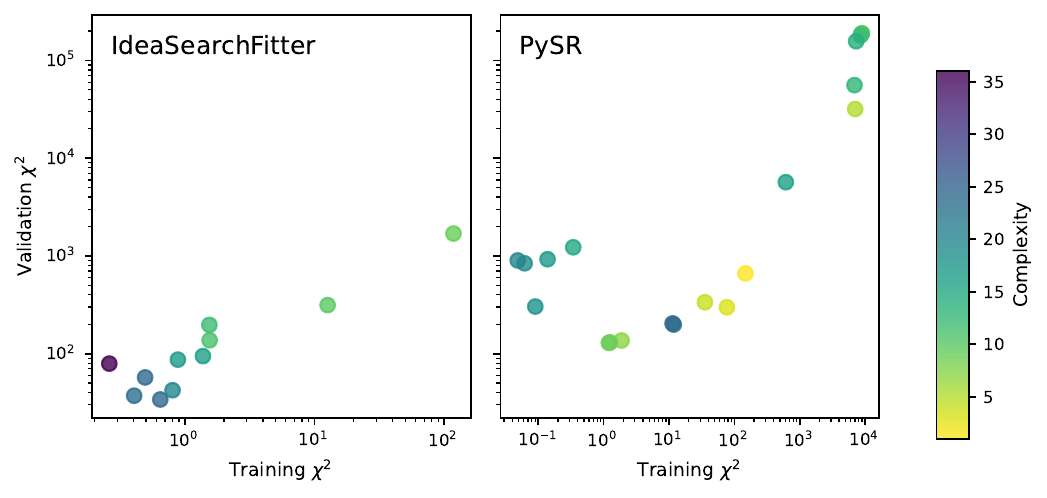}
\caption{Training–validation $\chi^2$ comparison for the anti-up quark ($\bar{u}$) Pareto front. Axes are logarithmic; colour encodes expression complexity (node count). Left: IdeaSearchFitter exhibits a coherent frontier where training and validation $\chi^2$ decrease jointly as complexity increases. Right: PySR displays two regimes. In the low‑training‑error regime (left portion of the panel), increasing complexity reduces training $\chi^2$ while raising validation $\chi^2$. In the higher‑training‑error regime (right portion), training and validation $\chi^2$ decrease together, but complexity changes are erratic, signaling poor numerical stability. The best validation point of IdeaSearchFitter attains a smaller validation $\chi^2$ than PySR, and its associated training $\chi^2$ is at approximately the same level as the elbow of PySR’s frontier.}
\label{fig:pareto_ubar}
\end{figure}

\textbf{Training–validation $\chi^2$ Pareto fronts.}
For each flavour, we construct the Pareto frontier in the plane spanned by training $\chi^2$ and validation $\chi^2$, with model complexity encoded by colour. Fig.~\ref{fig:pareto_ubar} shows representative results for the anti-up quark ($\bar{u}$). As shown in Fig.~\ref{fig:pareto_ubar}, the IdeaSearchFitter frontier displays coordinated reductions in training and validation $\chi^2$ with increasing, smoothly varying complexity, and no indications of numerical instability. By contrast, PySR exhibits a pronounced overfitting regime where validation $\chi^2$ rises as complexity is increased to drive down training $\chi^2$, followed by a joint decrease accompanied by irregular complexity jumps. At comparable training fit strength (near PySR’s elbow), IdeaSearchFitter attains a strictly lower validation $\chi^2$, demonstrating good extrapolation at comparable in‑domain fit strength.

\begin{figure}[ht!]
\centering
\begin{subfigure}[b]{.45\linewidth}
\includegraphics[width=\linewidth]{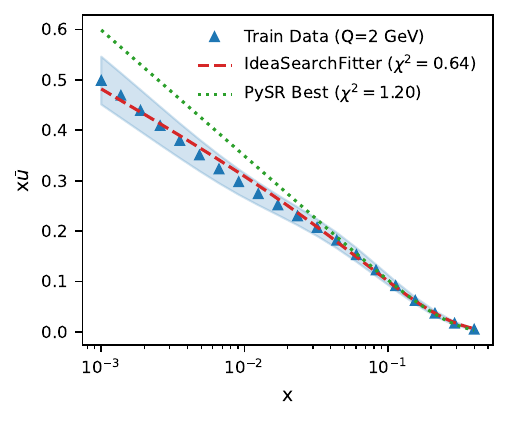}
\caption{$\bar{u}$ at $Q=2\,\mathrm{GeV}$}
\end{subfigure}\hfill
\begin{subfigure}[b]{.45\linewidth}
\includegraphics[width=\linewidth]{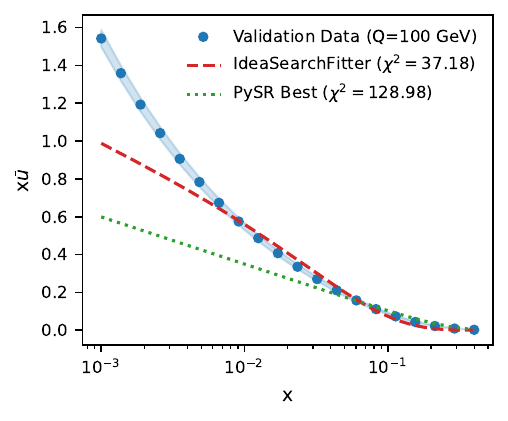}
\caption{$\bar{u}$ at $Q=100\,\mathrm{GeV}$}
\end{subfigure}
\caption{In‑domain and out-of-domain comparison for $\bar{u}$ at $Q=2$ and $100\,\mathrm{GeV}$. IdeaSearchFitter and PySR reach comparable accuracy within uncertainties for in-domain cases, while IdeaSearchFitter works better for out-of-domain cases.(c.l.=68\%)}
\label{fig:ubar_train}
\end{figure}

\textbf{Best validation models: in‑domain parity and strong out‑of‑domain performance.}
To isolate extrapolation effects, we select, for each method, the model with the lowest validation $\chi^2$ and compare against data at representative in‑domain scales ($Q=2\,\mathrm{GeV}$) and the out‑of‑domain target ($Q=100\,\mathrm{GeV}$). Fig.~\ref{fig:ubar_train} shows that in‑domain accuracy is comparable between the two methods—each recovers the broad $x$‑dependence within uncertainties at $Q=2\,\mathrm{GeV}$. However, at $Q=100\,\mathrm{GeV}$, IdeaSearchFitter remains within the reference uncertainty band better and achieves markedly lower validation $\chi^2$ ($\bar{u}$: $37.19$ vs.\ $128.98$), while PySR deviates substantially due to extrapolation‑unstable functional nests.

For the anti-up quark ($\bar{u}$), a compact ansatz grounded in DGLAP logarithmic evolution and Regge‑motivated small‑$x$ growth found by IdeaSearchFitter is
\begin{equation}
\label{eq:ubar_ansatz}
xf_{\bar{u}}(x,Q)\;=\;p_1\,\log\!\big(1+ p_2\log(Q)\big)\,\log\!\Big(\tfrac{1}{x}\Big)\,\big(1-x\big)^{\,p_4 + p_5\,\log Q},
\end{equation}

Most of the expressions found by IdeaSearchFitter are interpretable, avoid deep nesting, and remain numerically well‑defined over the evaluation domain, which directly translates into extrapolation stability.

\textbf{Summary across flavours: semantic, non‑nested forms yield stability.}
Tab.~\ref{tab:2d_models} reports, for each flavour, the best validation model from IdeaSearchFitter and PySR. IdeaSearchFitter consistently returns non‑nested, domain‑coherent functional forms with few parameters and moderate complexity. These ansatz avoid catastrophic numerical behaviour in extrapolation (no NaN/Inf, no discontinuities) and achieve substantially lower validation $\chi^2$ in all flavours, despite PySR occasionally attaining lower training $\chi^2$ through high‑variance nests. The results corroborate that semantic guidance, via natural‑language rationales encoding QCD priors and phenomenological insights, translates into quantitatively better extrapolation on PDF data.

\begin{table}[t]
\centering
\caption{\textbf{Optimal validation models per flavor($xf_i(x,Q)$).} 
IdeaSearchFitter achieves lower $\chi^2$ at matched complexity, 
with interpretable, stable forms.}
\label{tab:2d_models}
\resizebox{\textwidth}{!}{%
\begin{tabular}{llXcccrr}
\toprule
Flavor & Method & Ansatz Sketch & Num.\ Params.\ & Complexity & Train $\chi^2$ & Min. Val.\ $\chi^2$ \\
\midrule
$\mathrm{d}_\mathrm{v}$
& IdeaSearchFitter
& $p_0 \sqrt{x}(1-x)^{p_1\tanh\!\bigl(\arctan(\log Q)\bigr)}$
& \textbf{2} & 15 & \textbf{5.56} & \textbf{20.90} \\
& PySR
& $\sin\!\left(-\dfrac{p_0}{\sqrt{Q}}+\sqrt{\dfrac{x}{p_1}+p_2}\right)\,p_3$
& 4 & 14 & 0.49 & 40.42 \\
\midrule
$\bar{\mathrm{u}}$
& IdeaSearchFitter
& $p_0\,\log\!\big(1+ p_1\log(Q)\big)\,\log\!\Big(\tfrac{1}{x}\Big)\,\big(1-x\big)^{\,p_2 + p_3\,\log Q}$
& 4& 24& \textbf{0.64}& \textbf{34.00}\\
& PySR
& $\log\!\left(x+\dfrac{p_0}{\sin x}\right)\,p_1$
& \textbf{2} & 9 & 1.20 & 128.98 \\
\midrule
$\bar{\mathrm{d}}$
& IdeaSearchFitter
& $p_0\exp\!\left(\dfrac{p_1\sqrt{\log Q}}{\log(p_2/x)}\right)\log Q$
& 3 & 17 & \textbf{0.58} & \textbf{26.77} \\
& PySR
& $(x-p_0)^2$
& \textbf{1} & 4 & 14.84 & 207.84 \\
\midrule
$\mathrm{s}$
& IdeaSearchFitter
& $p_0\log(1/x)\tanh(\log Q)\exp(-p_1 x)$
& 2 & 16 & \textbf{0.23} & \textbf{12.77} \\
& PySR
& $(\log x)^2\,p_0-p_1$
& 2 & 7 & 0.21 & 17.07 \\
\midrule
$\mathrm{g}$
& IdeaSearchFitter
& $\log Q\log(1/x)\bigl(1-\sqrt{x}\bigr)^{p_0\log Q}$
& \textbf{1} & 17 & \textbf{2.83} & \textbf{117.99} \\
& PySR
& $\frac{p_0}{\frac{p_1}{Q-p_2}
+[-\sin(p_3 x) + p_4/Q]^2+x+\sqrt{x}}
-p_5 x$
& 6 & 26 & 109.26 & 164.99 \\
\midrule
$\mathrm{u}_\mathrm{v}$
& IdeaSearchFitter
& $p_0 x^{p_1}\exp\!\bigl(-p_2 x\arctan(\sqrt{\log Q})\bigr)$
& \textbf{3} & 16 & \textbf{6.35} & \textbf{65.64} \\
& PySR
& $\sin\!\left((\sqrt{x}+p_0)\cdot p_1\right)(x+p_2)+\dfrac{x - p_3}{p_4 Q}$
& 5 & 19 & 0.35 & 97.00 \\
\bottomrule
\end{tabular}%
}
\end{table}

The complete set of Pareto fronts and scale‑resolved comparisons for all flavours, including $Q=2,5,100\,\mathrm{GeV}$ panels, are provided in Appendix~\ref{app:full_2d}.

\section{Conclusion}
\label{sec:conclusion}

In this work, we sought to address the fundamental challenge in SR: the tension between predictive accuracy and model interpretability, which arises from navigating a vast, combinatorially complex search space. We introduced IdeaSearchFitter, a novel framework that reframes this challenge by shifting the search paradigm from a purely \emph{syntactic} exploration of expression trees to a \emph{semantic} search over an ``interpretable ansatz space". At the heart of our approach is an iterated agent powered by LLMs, which act as reasoning engines to generate candidate hypotheses. This ``explain-then-formalize" workflow, where LLMs propose natural-language rationales before symbolic forms, inherently constrains the search to physically plausible and conceptually coherent models, mitigating overfitting and enhancing scientific utility.

Our empirical validation demonstrated the effectiveness of this semantic approach across multiple domains. On the FSReD, IdeaSearchFitter achieved competitive performance, significantly outperforming established baselines, such as PySR and AI-Feynman, achieving recovery rates of 82.5\% in noise-free conditions and maintaining strong robustness with a 71.7\% recovery rate at high noise levels ($\gamma=0.1$). 

The practical utility of IdeaSearchFitter was underscored in its application to real-world and frontier scientific problems. On PMLB datasets, it discovered mechanistically aligned models, such as logarithmic memory effects in agricultural yields and regime-separated dynamics in supernova light curves, that achieved a good accuracy-complexity trade-off without the overfitting artifacts common to baseline methods. In a challenging high-energy physics case study, the framework derived compact, interpretable parametrizations for PDFs that not only fit data with good $\chi^2/\mathrm{ndf}$ but also exhibited promising extrapolation capabilities in our tests, a critical requirement for physical models. This result highlights the framework's potential to contribute tangible insights in domains where interpretability and physical fidelity are paramount.

Looking forward, our work opens several promising avenues. The framework's modular design invites the creation of hybrid systems that combine the semantic guidance of LLMs with the efficiency of traditional syntactic operators, potentially yielding the best of both worlds. Furthermore, the intelligent orchestration of diverse LLM agents, each with potentially different strengths, presents another avenue for enhancing search efficiency and robustness. While our current implementation relies on the quality and availability of LLM APIs, future work could explore the integration of smaller, fine-tuned, or open-source models to reduce costs, improve accessibility, and ensure data security, particularly for sensitive applications. Ultimately, the IdeaSearch framework provides a promising and generalizable blueprint for agentic AI in science, positioning LLM-driven semantic search as a key methodology for accelerating data-driven discovery not only in symbolic regression but across the scientific disciplines.

\section*{Acknowledgments}
This work is supported by the National Science Foundation of China under contract No.
12425505, 12235001, 12175016, U2230402, the Fundamental Research Funds for the Central Universities, Peking University, and the Fundamental Research Funds for the Central Universities, Beijing Normal University
\bibliographystyle{unsrtnat}
\bibliography{cite}

\appendix

\section{Formalism of an Iterated Ideal Agent}
\label{app:agent_formalism}

This appendix provides a rigorous mathematical framework for the iterated discovery process and the search operator $\mathcal{T}$ introduced in Section~\ref{sec:methodology}. The goal is to establish a formal language to precisely distinguish between different search paradigms, such as the syntactic search common in genetic programming and the semantic search enabled by our framework. A practical LLM-driven system like IdeaSearchFitter can be viewed as an ensemble of agents operating under this formalism.

\begin{definition}[Discovery as Search in a Functional Space]
Many scientific discovery problems of interest can be usefully modelled as searches for hypotheses in an appropriately structured function space. A hypothesis is represented as a function, e.g., $f(x_1, \dots, x_n)$, that models the relationship between variables. The space contains all possible hypotheses, and the norm, $|f|$, can be defined to quantify a desired property, such as fitting error (e.g., $\chi^2$) or model complexity.
\end{definition}

\paragraph{Motivation.} This abstraction allows us to apply concepts from analysis to the problem of discovery. By treating hypotheses as points in such space, the search for a good model becomes equivalent to navigating this space to find a region of "high value" (e.g., low error and low complexity). Remark: by ‘space’ we mean a function space equipped with appropriate structure together with one or more evaluation functionals. We do not require the evaluation functional to be a mathematical norm in the strict sense; hence below we use the neutral term ‘quality/measure’ where appropriate.

\begin{definition}[The Agent Operator]
An agent operator $\mathcal{T}$ is a mapping from a set of input functions (existing hypotheses) in a normed functional space $\mathrm{C}_1$ to a set of output functions (new hypotheses) in another normed functional space $\mathrm{C}_2$:
\[
\mathcal{T}: \{f | f\in\mathrm{C}_1\} \to \{g | g\in \mathrm{C}_2\}.
\]
\end{definition}

\paragraph{Motivation.} The operator $\mathcal{T}$ represents the core "reasoning engine" of the discovery agent. In genetic programming, $\mathcal{T}$ would be the application of mutation and crossover rules. In our framework, $\mathcal{T}$ corresponds to the process of an LLM generating new symbolic expressions based on a prompt containing prior successful hypotheses.

\begin{definition}[Capability and the Guiding Measure]
For a fixed input set of hypotheses ${f} \subseteq \mathrm{C}_1$, the capability of an agent $\mathcal{T}$, denoted $\gamma(\mathcal{T})$, is the (normalized) measure of its output set:
\[
\gamma(\mathcal{T}) := \mu(\mathcal{T}({f})).
\]
\end{definition}

\paragraph{Motivation.} The choice of the measure $\mu$ is critical, as it encodes the values and biases of the search. This is where the distinction between syntactic and semantic search is formally captured:
\begin{itemize}
\item \textbf{A Syntactic Measure ($\mu_{\text{syn}}$)} assigns a high value to functions based on their structural properties. For example, it might favor expressions with fewer nodes (simplicity) or those that achieve a lower training error, irrespective of their physical meaning. This guides the search within a purely \textit{syntactic space}.
\item \textbf{A Semantic Measure ($\mu_{\text{sem}}$)} assigns a high value to functions based on their conceptual and domain-specific coherence. For example, it would favor expressions that satisfy physical priors (e.g., dimensional homogeneity, conservation laws, correct boundary conditions) or are supported by a plausible natural-language rationale. This guides the search within a more constrained \textit{semantic space}.
In practice constructing a faithful $\mu_{\text{sem}}$ is challenging—it depends on the available domain priors, the fidelity of natural‑language rationales, and the limitations of the agent. LLMs can approximate $\mu_{\text{sem}}$ under favourable prompting and data regimes, but they do not guarantee discovery of physically correct hypotheses without expert oversight.
\end{itemize}

Here $\mu$ denotes a generic nonnegative valuation functional (or measure) on sets of hypotheses. We keep $\mu$ intentionally abstract to encompass different possible choices.

\begin{definition}[The Iterated Agent]
An agent is an \textit{iterated} agent if its output space is a subset of its input space ($\mathrm{C}_2 \subseteq \mathrm{C}_1$). This closure property ensures that the outputs of the operator can be fed back as inputs for subsequent steps of refinement.
\end{definition}

\begin{definition}[Search Trajectory]
For a single-input iterated agent (where the input set contains one hypothesis $f_i$), a search trajectory is a sequence of hypotheses $f_0 \to f_1 \to f_2 \to \cdots$, such that for each step $i \ge 0$:
\[
f_{i+1} \in \mathcal{T}(f_i).
\]
\end{definition}

\paragraph{Motivation.} This sequence represents a single, coherent line of inquiry or refinement. The initial hypothesis, $f_0$, can be a null model or a simple guess. The agent then iteratively applies its operator $\mathcal{T}$ to generate a path through the hypothesis space. The effectiveness of the agent depends on its ability to guide this trajectory towards regions of high measure $\mu$. In a multi-island system like IdeaSearchFitter, multiple such trajectories are explored in parallel, allowing for a broad yet guided exploration of the search space.

to apply this formalism in experiments, one must at least fix the choice of functional space and the operator $\mathcal{T}$; these implementation choices are described for IdeaSearchFitter in App.~\ref{app:ideasearch_impl} and the public code repository\cite{ideasearchfitter_repo}.

\section{Implementation Details of the IdeaSearchFitter Evolutionary Search}
\label{app:ideasearch_impl}

This appendix provides comprehensive technical details supporting the evolutionary search architecture of IdeaSearchFitter, as described in Section~\ref{sec:methodology}. It includes pseudocode sketches, algorithmic formulas, the two-stage ansatz agent protocol, repository schema, and Pareto reporting procedures. Full prompt templates, hyperparameter specifications, and reproducible implementation scripts are archived in the open-source release accompanying this paper~\cite{ideasearchfitter_repo}. These details ensure transparency and facilitate replication across diverse scientific domains.

\subsection{Preprocessing}
The pipeline consumes two complementary information streams and fuses them into a persistent, enriched problem context used throughout the search. First, it processes structured metadata that includes variable names, optional units, brief natural-language descriptions, and any explicit domain priors provided by the user. These inputs are canonicalized and then expanded via a single LLM analysis to yield an enriched, internally consistent context together with suggested heuristics for downstream generation. Second, it ingests numerical data and associated uncertainty information—observed samples with measurement uncertainties or weights—and computes simple diagnostics such as variance, approximate signal-to-noise ratio, and outlier flags. This numerical stream is automatically inspected to produce lightweight reports, which are attached to the enriched context.

These two streams are then fused into a single contextual prompt that is supplied to the generation agents at each epoch. In addition, a snapshot of the current top-performing ideas from the central repository is appended to provide explicit, high-quality seeds for the LLMs.

\textbf{Inputs:} User-supplied metadata (variable names, optional units, brief natural-language description) and numerical dataset (input features, target values, optional uncertainties or weights).

\textbf{Steps:}
\begin{enumerate}
    \item Canonicalize variable names and units.
    \item Compute diagnostic statistics: mean, variance and outlier flags.
    \item Perform a single LLM analysis pass on the metadata to expand the description into an enriched context (3-5 sentences), incorporating domain priors and suggested primitives. Persist this as the core problem prompt, appended with current top-performing ideas from the repository.
\end{enumerate}

This enriched context is supplied to generation agents at each epoch, promoting semantically coherent proposals.

\subsection{Two-Stage Ansatz Agent}
Each candidate ansatz is generated via a two-stage protocol~\cite{stepbystep}, decoupling semantic ideation from syntactic canonicalization to enhance diversity while ensuring parseability.

\textbf{Pseudocode:}
\begin{verbatim}
def generate_ansatz(context, top_K_ideas, seed):
    # Stage 1: Proposal Generation
    # Higher temperature for creativity
    prompt = build_prompt(context, top_K_ideas, seed)
    rationale, sketches = llm.generate(prompt, model=distributor.select())
    
    # Stage 2: Ansatz Extraction
    for sketch in sketches:
        # Lower temperature, constrained
        canonical_expr, metadata = llm.extract(sketch, rationale)
        if parse_and_validate(canonical_expr):  # Bounded retries on failure
            return canonical_expr, metadata  # Includes param names, bounds, units
    return None  # Reject on exhaustion
\end{verbatim}

The proposal stage yields concise rationales and sketch expressions. The extraction stage outputs a single, parseable symbolic expression in numexpr-compatible syntax, with metadata for downstream fitting.

\subsection{Distributor for LLM Allocation}
The distributor employs a softmax policy to allocate generation requests across an ensemble of LLM backends, adapting online to performance.

Let $s_i$ denote the exponential moving average score for backend $i$ (updated via recent proposal success rates). With fixed temperature $T_{\text{model}}$, the selection probability is:
\begin{equation}
p_i = \frac{\exp(s_i / T)}{\sum_j \exp(s_j / T)}.
\end{equation}

This bandit-like mechanism could balance exploitation and exploration (Section~\ref{sec:benchmarks}).

\subsection{Multi-Island Evolutionary Lifecycle}
\begin{figure}
\centering
\includegraphics[width=\linewidth]{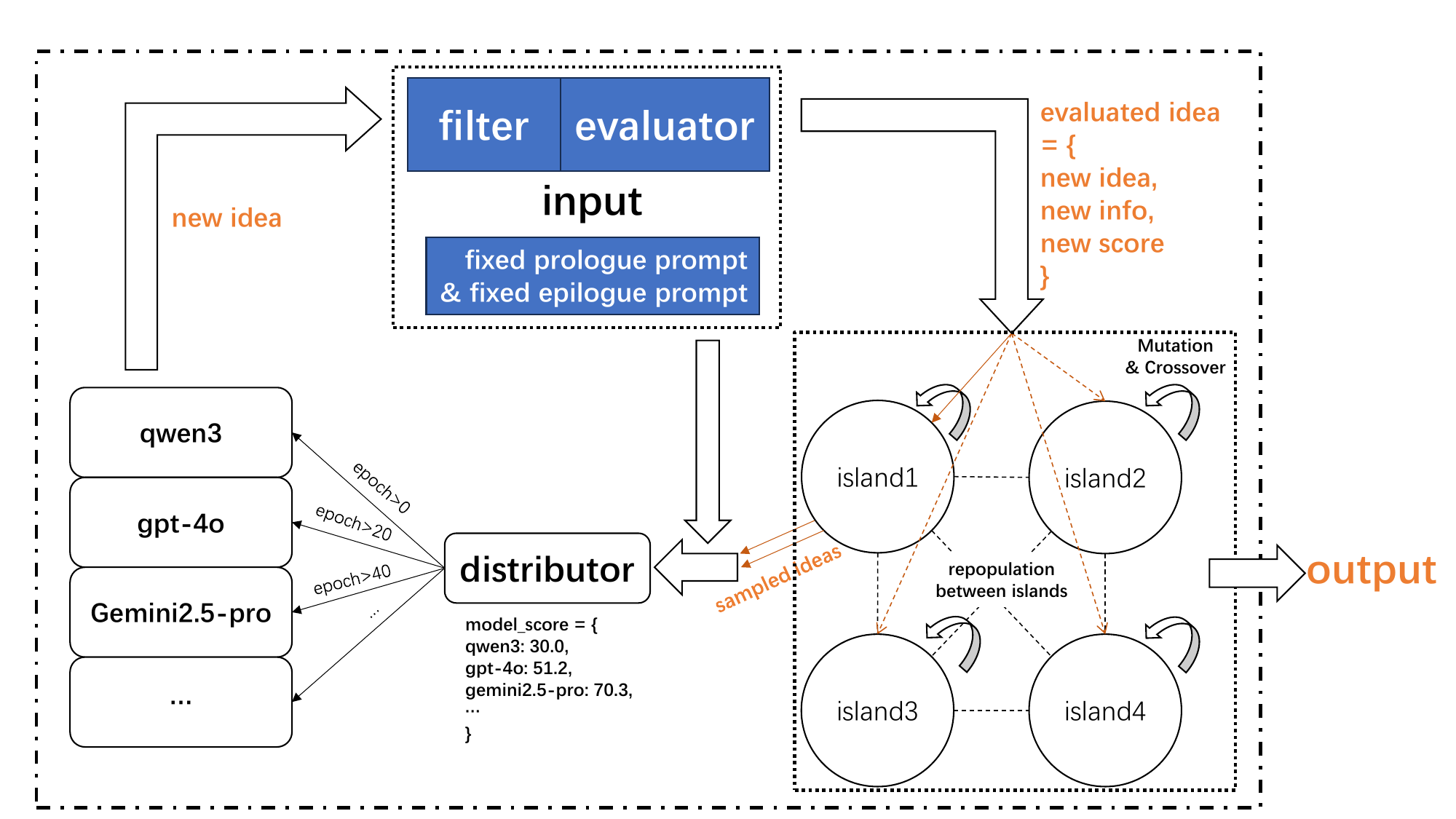}
\caption{\textbf{Schematic of the inner search loop in IdeaSearchFitter.} The left panel depicts an ensemble of LLMs (e.g., Qwen3, GPT-4o, Gemini 1.5-Pro) prompted with a fixed prologue/epilogue template and the filtered data description (new idea input). An epoch-aware distributor allocates generation requests across models via a multi-armed bandit strategy, guided by learned model scores. Generated proposals (new ideas) undergo a lightweight filter and statistical evaluation (e.g., reduced $\chi^2/\mathrm{ndf}$), yielding evaluated idea objects ${ \text{idea}, \text{info}, \text{score} }$ stored in a centralized repository. The right panel illustrates sampling of evaluated ideas into a multi-island evolutionary population (four islands shown), where local genetic operators (mutation and crossover, orange arrows) refine subpopulations, and inter-island repopulation (dashed arrows) promotes diversity and averts premature convergence. Iterative cycles of LLM generation, evaluation, and evolutionary updates progressively enhance hypothesis quality, culminating in candidate sets for downstream Pareto analysis that balances fit quality against structural complexity (output).}
\label{fig:is flow chart}
\end{figure}

The search operates as a repopulation-only multi-island process (8-12 islands, per hyperparameters in Appendix~\ref{app:hyperparameters}), with LLM proposals as the primary variation source. Syntactic tree mutations and crossovers are implemented for hybrid use but disabled in the reported experiments to isolate semantic effects. The exact cycle is shown in Fig.~\ref{fig:is flow chart}.

\textbf{Epoch Flow:}
\begin{enumerate}
    \item Distribute per-island quotas via the distributor, conditioning on local population and local top ideas.
    \item Fit parameters, evaluate (reduced $\chi^2/\mathrm{ndf}$), and archive successful candidates.
    \item Every several epochs, repopulate the worst islands by sampling the repository with the global best ideas.
\end{enumerate}

This configuration sustains semantic lineages, preventing convergence too fast.

\subsection{Parameter Fitting, Evaluation, and Rejection}
Fitting uses bounded nonlinear least squares (L‑BFGS‑B via SciPy). By default we set coefficient bounds to [-10,10] as a heuristic; these bounds can and should be adjusted to problem scale or replaced by problem‑specific priors.

\textbf{Evaluation:} Primary metric is reduced $\chi^2$/ndf as introduced in Section~\ref{sec:methodology}.
If per‑sample uncertainties $\sigma_i$ are not provided, we use weightless MSE with explicit note in the archive.

\textbf{Rejection Rules:} Discard candidates with NaNs, divergences or expressions whose unit don't match the target (Optional).

\subsection{Repository Schema}
The central repository stores evaluated ideas with the following schema:

\begin{itemize}
    \item \texttt{expression}: Canonical string (numexpr format).
    \item \texttt{params}: Fitted values(JSON).
    \item \texttt{chi2/ndf}: Reduced $\chi^2$/ndf.
    \item \texttt{complexity}: Node count.
    \texttt{rationale}: Natural-language justification.
    \item \texttt{origin}: Metadata (LLM ID, epoch, island).
\end{itemize}

\subsection{Reproducibility and Artifacts}
Prompt templates (e.g., rationale-focused for generation, constrained for extraction), exact hyperparameters (e.g., epoch budgets, timeouts), and seeds are detailed in the repository~\cite{ideasearchfitter_repo}. The framework integrates with IdeaSearch~\cite{ideasearch2025} for modular extensions, released concurrently.

\section{Hyperparameters for Symbolic Regression Baselines and IdeaSearchFitter}
\label{app:hyperparameters}

To facilitate reproducibility, this appendix details the hyperparameter configurations used for the baseline SR methods (PySR~\cite{pysr}, Operon~\cite{operon}, and AI-Feynman~\cite{aifeynman}) as well as IdeaSearchFitter. These settings were optimized for computational fairness. Configurations for IdeaSearchFitter distinguish between "fast" (benchmark-oriented) and "expert" (application-oriented) modes.

The hyperparameters are summarized in Tab.~\ref{table:hyperparams2}. Note that seeds were randomized within bounds for stochastic elements, and operator sets were restricted to promote comparability while allowing domain-relevant expressivity. Where possible we tuned settings to produce comparable computational effort across methods; nevertheless, results can depend on implementation details, hardware, and provider‑specific factors.

\begin{table}[H]
    \vspace{-1em}
    \caption{Hyperparameter sets for SR baselines (part 2).}
    \label{table:hyperparams2}
    \def\arraystretch{1.2}
    \begin{center}
        \bgroup
        \small
        \setlength{\tabcolsep}{0.1em}
        \begin{tabular}{c|l} 
            \toprule
            \multicolumn{1}{c|}{\bf Method} & \multicolumn{1}{c}{\bf Hyperparameter sets} \\
            \midrule
            AI-Feynman
            & \emph{brute force limit time}: 60s, \\
            & \emph{symbols to be used in the brute force code}: \\
            & ['+-*/', 'cos', 'sin', 'tan', 'arc cos', 'arc sin', 'arc tan', 'squared', 'exponential', 'log', 'inverse'],
            \\
            & \emph{polyfit degree}: $number\,of\,variables-1$, \emph{neural network epochs}: 500, \emph{test percentage}: 20\%\\
            & \emph{Batch Size}: 2048, \emph{learning rate}: 0.01, \emph{weight decay}: 0.01\\
            \\
            PySR 
            & \emph{procs}: 5, \emph{populations}: 10, \emph{population\_size}: 40, \emph{ncyclesperiteration}: 500, \\
            & \emph{niterations}: 50000, \emph{timeout\_in\_seconds}: 7200, \emph{maxsize}: 50, \\
            & \emph{binary\_operators}: ['*', '+', '-', '/'], \emph{unary\_operators}: ['sin', 'cos', 'exp', 'log', 'square', 'sqrt'], \\
            & \emph{nested\_constraints}: \{sin: \{sin: 0, cos: 0\}, cos: \{sin: 0, cos: 0\}, exp: \{exp: 0\}, log: \{log: 0\}\}, \\
            & \emph{progress}: False, \emph{weight\_randomize}: 0.1, \emph{precision}: 32, \emph{warm\_start}: False, \\
            & \emph{turbo}: True, \emph{update}: False \\
            \\ 
            Operon 
            & \emph{generations}: 1000, \emph{max\_evaluations}: 500000, \emph{local\_iterations}: 5, \\
            & \emph{population\_size}: 500, \emph{pool\_size}: 500, \emph{p\_crossover}: 1.0, \\
            & \emph{p\_mutation}: 0.25, \emph{seed}: random(1,1000000),  \\
            & \emph{primitive\_set}: ['Constant', 'Variable', 'Add', 'Mul', 'Div'], \\
            & \emph{min\_length}: 1, \emph{max\_length}: 50, \emph{max\_depth}: 10, \\
            & \emph{selector}: 'TournamentSelector', \emph{tournament\_size}: 5, \\
            & \emph{crossover}: 'SubtreeCrossover(0.9, max\_depth=10, max\_length=50)', \\
            & \emph{mutations}: ['NormalOnePointMutation', 'ChangeVariableMutation', 'ChangeFunctionMutation', \\
            & 'ReplaceSubtreeMutation'] \\
            \\
            IdeaSearchFitter
            & \emph{models}: ['gpt-5-mini', 'gpt-5', 'qwen3', 'qwen-plus', 'gemini-2.5-flash',
            \\ & 'deepseek-v3', 'grok-4', 'doubao','gemini-2.5-pro(I)', 'gemini-2.5-pro(II)'], \\
            & \emph{fuzzy\_translator}: 'gemini-2.5-flash', \emph{shutdown\_score}: 79.9(fast)/89.9(expert), \emph{samplers\_num}: 3, \\
            & \emph{sample\_temperature}: 20.0, \emph{evaluators\_num}: 3, \emph{examples\_num}: 2(fast)/5(expert), \\
            & \emph{generate\_num}: 2, \emph{model\_assess\_average\_order}: 15.0, \\
            & \emph{model\_assess\_initial\_score}: 20.0, \emph{island\_num}: 8(fast)/12(expert), \emph{cycle\_num}: 5(fast)/30(expert), \\ & \emph{unit\_interaction\_num}: 6(fast)/10(expert), \\ & \emph{auto\_polish}:True(fast)/False(expert)
            \\ & \emph{constant\_whitelist}: ['1','2','pi'],
            \emph{constant\_whitelist}: \{'1':1,'2':2,'pi':np.pi\}, polish\_level: 3(fast)/16(expert) \\
            \\
            \bottomrule
        \end{tabular}
        \egroup
    \end{center}
\end{table}

\section{Characterizing LLM Reasoning Styles Through Dynamic Evaluation}
\label{app:llm_benchmark}

Standard LLM benchmarks measure performance through static, single-shot accuracy metrics. While these metrics establish a general capability ranking, they treat problem-solving as a monolithic task, obscuring the underlying process. Consequently, a model that finds a solution in one step is indistinguishable from one that reaches the same solution through a lengthy, iterative process of refinement. This limitation masks crucial differences in problem-solving strategies, which are fundamental to understanding model intelligence.

To overcome this, we conceptualize problem-solving through the lens of an \textbf{iterative agent}. Instead of producing a final answer in a single step, this agent generates a sequence of hypotheses through an iterative process. This structure is paramount, as it enables \textbf{step-by-step evaluation} and the use of more complex, fine-grained reward signals to guide the discovery process. \texttt{IdeaSearchFitter} serves as a concrete implementation of this dynamic paradigm. Within our framework, the LLM is the core of this iterative agent, driving the search by proposing new hypotheses at each epoch. By embedding different LLMs, we can analyze their entire solution trajectory ($f_0 \to f_1 \to \cdots \to f_{\text{true}}$). This allows us to characterize distinct "reasoning phenotypes"—such as rapid convergers versus persistent explorers—based on the properties of the path taken, a dimension of performance entirely invisible to static benchmarks.

\subsection{Experimental Setup}

We curated 120 challenging problems from the FSReD dataset and embedded ten prominent LLMs as the generative core of IdeaSearchFitter. Every model attempted the full test battery under identical framework settings (namely the “fast" configuration in Tab.~\ref{table:hyperparams2}, with the ensemble replaced by a single LLM). In addition to the final success rate, we recorded several complementary metrics:

\begin{itemize}
    \item \textbf{Iter@1, Iter@10, Iter@100:} The number of problems solved within 1, 10, and 100 epochs, respectively. These metrics capture both immediate problem-solving ability and sustained search stamina.

    \item \textbf{Median Time:} The typical wall-clock time per problem, accounting for API latency and inference speed.
\end{itemize}

\begin{figure}[ht!]
    \centering
    \includegraphics[width=0.7\linewidth]{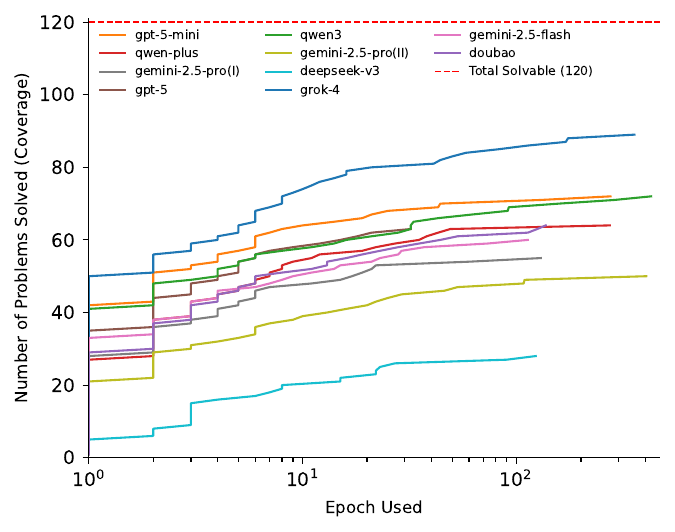}
    \caption{\textbf{Cumulative solved problems versus evolutionary epochs for ten LLMs on 120-problem FSReD ($\gamma=0.1$ noise).} Lines trace unique recoveries (log-scale epochs $1$--$400$). Diverse phenotypes emerge: rapid convergers (e.g., gpt-5-mini, steep initial rise) versus persistent explorers (e.g., qwen3, sustained growth to 60 solves; cf. Tab.~\ref{tab:llm_comparison}).}
    \label{fig:coverage_vs_epoch}
\end{figure}

\begin{figure}[ht!]
    \centering
    \includegraphics[width=\linewidth]{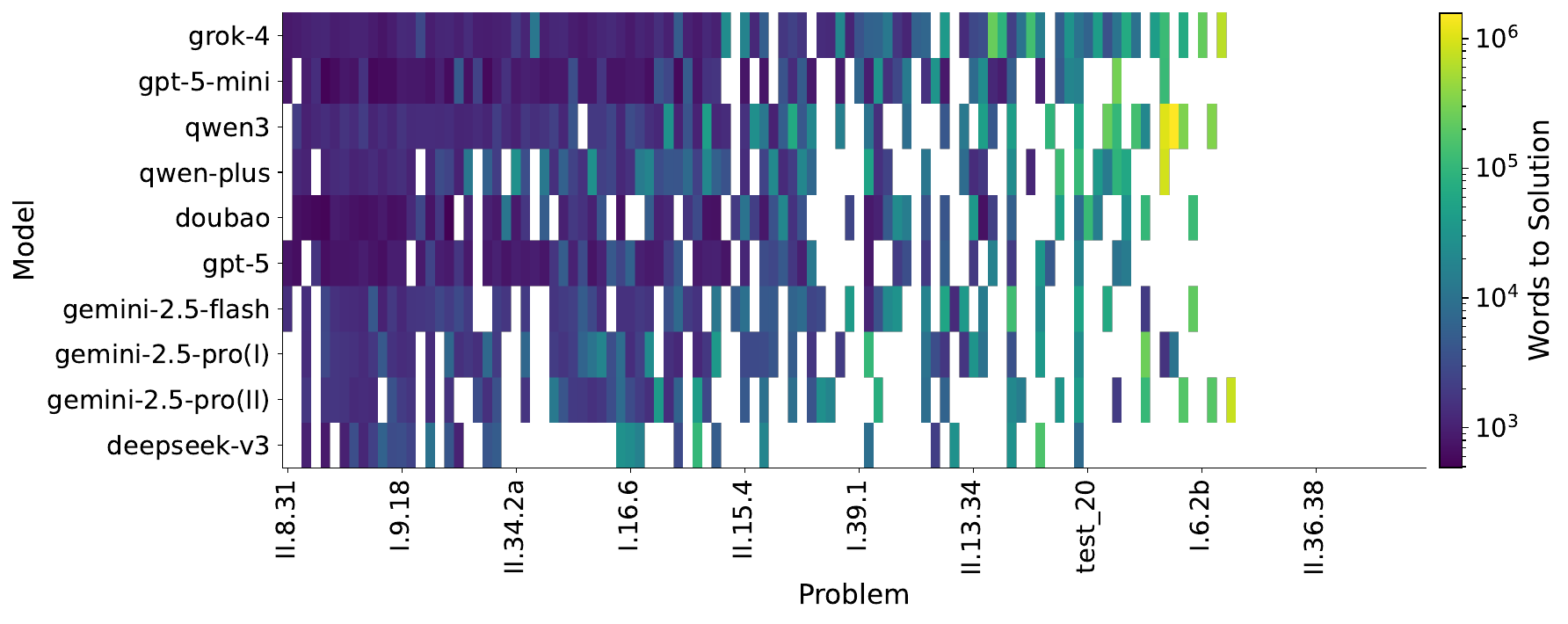}
    \caption{\textbf{Heatmap of word consumption for successful SR solves across LLMs and problems.} The matrix visualizes the words' usage by ten LLMs (rows: grok-4, gpt-5-mini, qwen3, qwen-plus, doubao, gpt-5, gemini-2.5-flash, gemini-2.5-pro(I), gemini-2.5-pro(II), deepseek-v3) to recover ground-truth expressions on 120 challenging FSReD problems ($\gamma=0.1$ noise; columns ordered by increasing average difficulty). Cell colors encode the total words used for successful solves (lighter shades indicate higher consumption, up to $10^6$; white cells denote failures). Highlighting the value of multi-agent ensembles for robust discovery, as no model achieves uniform coverage.}
    \label{fig:epochs_to_solution}
\end{figure}

\begin{figure}
    \centering
    \includegraphics[width=0.7\linewidth]{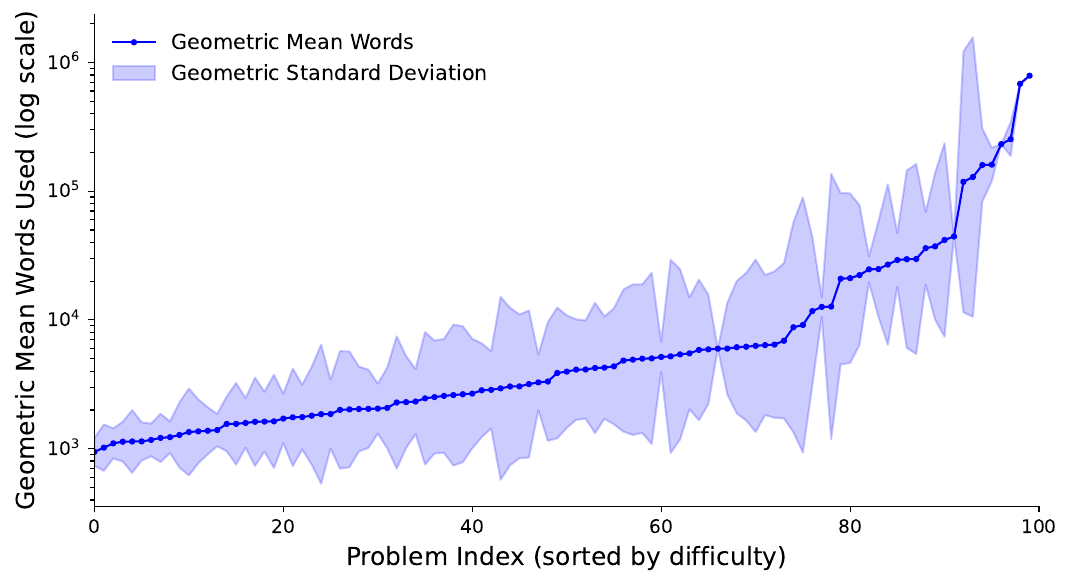}
    \caption{\textbf{Geometric mean words count required per problem across successful solves.} Displays words consumed by LLMs for 100 representative FSReD problems ($\gamma=0.1$ noise). Blue line: mean words used (log-scale $10^3$--$10^6$); blue shade: geometric SD. Highlights reasoning variability, motivating phenotype ensembles (cf. Tab.~\ref{tab:llm_comparison}, Fig.~\ref{fig:coverage_vs_epoch}).}
    \label{fig:words used}
\end{figure}

\begin{table}[ht!]
    \centering
    \caption{\textbf{Performance comparison of ten LLMs embedded in IdeaSearchFitter on a challenging 120-problem FSReD ($\gamma=0.1$ noise).} Ranks by solved problems; includes success (\%), median time (s), epochs (median, 3/4 quartile), median words used, and iter@k (solved within $k$ epochs).}
    \label{tab:llm_comparison}
    \resizebox{\textwidth}{!}{%
    \begin{tabular}{lccccccccc}
        \toprule
        \textbf{Model} & \textbf{Solved} & \textbf{Success (\%)} & \textbf{Median Time (s)}& \textbf{Median Epochs}& \textbf{3/4 Quartile Epochs}& \textbf{Median Words}& \textbf{Iter@1}& \textbf{Iter@10}& \textbf{Iter@100} \\
        \midrule
        \textbf{grok-4}& \textbf{89}& \textbf{74.17}& \textbf{200.00}& \textbf{2.00}& \textbf{7.00}& \textbf{1361.00}& \textbf{50}& \textbf{74}& \textbf{85}
\\
        \textbf{gpt-5-mini}& \textbf{72}& \textbf{60.00}& \textbf{193.00}& \textbf{2.00}& \textbf{5.00}& \textbf{938.00}& \textbf{42}& \textbf{64}& \textbf{70}
\\
        qwen3& 72& 60.00& 196.50& 2.00& 6.25& 1870.00& 41& 57& 69
\\
        qwen-plus& 64& 53.33& 163.50& 3.00& 7.00& 2612.50& 27& 54& 63
\\
        doubao& 64& 53.33& 229.00& 3.00& 7.00& 1640.00& 29& 51& 61
\\
        gpt-5& 63& 52.50& 172.00& 2.00& 4.00& 1032.00& 35& 58& 63
\\
        gemini-2.5-flash& 60& 50.00& 238.00& 2.00& 5.00& 2448.00& 33& 50& 59
\\
        gemini-2.5-pro(I)& 55& 45.83& 198.00& 2.00& 5.50& 2235.00& 28& 47& 54
\\
        gemini-2.5-pro(II)& 50& 41.67& 209.00& 3.00& 9.50& 3437.00& 21& 39& 47
\\
        deepseek-v3& 28& 23.33& 230.00& 4.00& 16.00& 4673.00& 5& 20& 27\\
        \bottomrule
    \end{tabular}%
    }
\end{table}

\subsection{Results and Analysis}

Tab.~\ref{tab:llm_comparison} presents the comprehensive results. The overall success rate ranking broadly aligns with general LLM leaderboards. This alignment confirms that our framework provides a valid measure of model capability. However, the process metrics reveal striking differences in reasoning styles.

Fig.~\ref{fig:coverage_vs_epoch} shows how problem coverage evolves with search epochs. All models show monotonic improvement, but their trajectories differ significantly. Some models solve most of their accessible problems quickly (steep initial curve). Others show sustained growth across many epochs (gradual climb).

Fig.~\ref{fig:epochs_to_solution} visualizes the word count required for each model to solve each problem. The distribution varies widely across models and problems. Some models consistently use concise reasoning (low word counts). Others require extensive deliberation (high word counts). Importantly, no single model solves all problems, demonstrating complementary strengths.

Fig.~\ref{fig:words used} reports the geometric-mean word budget expended by IdeaSearchFitter to recover a ground-truth expression for each of 100 solvable problems from the FSReD using IdeaSearchFitter (ordered by increasing average word count). The required consumption spans more than three orders of magnitude, from $10^{3}$ to $10^{6}$ words, establishing a broad cost spectrum that enables high-resolution differentiation among models.

For comparison, widely used static benchmarks (MATH-500~\cite{math500}, GPQA~\cite{GPQA}, AIME-2024~\cite{huggingface2024aime}, and PHYBench~\cite{phybench}) contain approximately $10^4$ words on average, substantially less than our dataset.

IdeaSearchFitter's extended dynamic range thus furnishes a more nuanced cost-success landscape, eschewing the artificial truncation of difficult problems inherent in static evaluations. This facilitates precise phenotyping of model capabilities, while maintaining efficiency on simpler instances.

\subsection{Identifying Distinct Reasoning Phenotypes}

Based on the data, we identify two prominent reasoning phenotypes:

\paragraph{The Rapid Converger.} This phenotype achieves high Iter@1 scores with tight epoch distributions. GPT-5-mini exemplifies this style. It solves 42 problems in the first epoch (Iter@1 = 42) and achieves most of its final coverage by epoch 10 (Iter@10 = 64 vs. Iter@100 = 70). Its 3/4 quartile epoch count is only 5.00, meaning 75\% of solved problems require at most 5 epochs. The median word count is 938, the lowest among competitive models. This phenotype excels at quickly identifying straightforward solutions with minimal reasoning overhead.

\paragraph{The Persistent Explorer.} This phenotype demonstrates sustained problem-solving ability across many epochs. Grok-4 exemplifies this style. While its median epoch is low (2.0), its 3/4 quartile is 7.0, a large gap indicating it continues solving difficult problems in later epochs. The coverage curve in Fig.~\ref{fig:coverage_vs_epoch} shows sustained growth. Grok-4 solves 50 problems by epoch 1, 74 by epoch 10, and reaches 85 by epoch 100, the highest coverage. This phenotype contributes both quick wins and long-term persistence.

Between these extremes lie intermediate phenotypes. For example, Gemini-2.5-Flash shows moderate Iter@1 (33) and moderate persistence (Iter@10 = 50, Iter@100 = 59). DeepSeek-v3 exhibits low immediate success (Iter@1 = 5) but gains substantially with more epochs, though its final coverage (28) remains limited.

\subsection{Implications for LLM-Driven Discovery}

This analysis demonstrates that IdeaSearchFitter enables a new form of LLM evaluation. Instead of a single accuracy score, we obtain a multi-dimensional profile of problem-solving style. The metrics capture speed (Iter@1), persistence (epoch distribution), efficiency (word count), and ultimate capability (final success rate).

The diversity of phenotypes has practical implications. No single model dominates across all metrics. An ensemble approach could leverage complementary strengths. For example, a Rapid Converger could quickly identify promising solution candidates, while a Persistent Explorer could tackle problems that require more extensive search. Our framework provides the methodology to identify these complementary capabilities and potentially orchestrate them effectively.

Beyond benchmark performance, this Appendix could contribute a methodology for understanding how different LLMs approach complex reasoning tasks. The dynamic metrics reveal qualitative differences that static accuracy scores cannot capture.

\section{Dataset Selection Pipeline from PMLB}
\label{app:dataset_selection}

To increase transparency and repeatability, we applied a conservative three‑stage filtering pipeline to curate a benchmark suite from the PMLB~\cite{pmlb}. The objective was to select observational datasets that represent genuine scientific discovery challenges, characterized by the absence of known ground-truth formulas and the potential for uncovering concise, interpretable symbolic laws. This appendix details the methodology, criteria, and outcomes of the process. We emphasize that these choices encode specific trade‑offs and may bias the resulting collection toward datasets amenable to concise symbolic descriptions.

\subsection{Stage 1: Rule-Based Filtering}

The initial stage applied deterministic rules to exclude datasets unsuitable for our evaluation goals, serving as a coarse filter to reduce the PMLB pool (initially over 400 regression tasks) to a manageable set of candidates. The exclusion criteria were as follows:

\begin{enumerate}[label=\alph*)]
    \item \textbf{Known Benchmark Sets:} Datasets containing keywords "feynman" or "strogatz" in their names were removed, as these are standard SR benchmarks with explicit ground-truth equations.
    
    \item \textbf{Classification Tasks:} Datasets were excluded if the target variable $y$ was non-numeric or exhibited low cardinality (fewer than five unique values), serving as a heuristic for categorical or ordinal targets incompatible with regression.
    
    \item \textbf{Dimensionality and Sample Size Constraints:} To prioritize problems amenable to interpretable symbolic modeling, we excluded datasets with more than 15 input features or fewer than 50 data points, ensuring sufficient samples for robust fitting and validation while maintaining low dimensionality.
\end{enumerate}

This stage yielded approximately 50 candidate datasets for further scrutiny.

\subsection{Stage 2: LLM-Assisted Semantic Filtering}

Surviving datasets underwent semantic analysis of their metadata using an LLM. For each dataset, the LLM was prompted with its descriptive metadata to extract properties such as problem context, variable meanings, and any indications of ground-truth formulas. Exclusion occurred if:
\begin{itemize}
    \item The metadata lacked a meaningful natural-language description of the scientific context or variables.
    \item An explicit ground-truth symbolic formula was present in the description.
\end{itemize}

The LLM also assigned scores for "description completeness" (clarity and detail of metadata) and "symbolic potential" (likelihood of an underlying concise law based on domain heuristics). These scores ranked candidates, prioritizing those with high potential for the manual review stage. This reduced the pool to about 20 datasets.
\paragraph{Caveat:} The LLM‑based scoring and metadata interpretation were used as heuristic aids rather than authoritative judgments. LLM outputs are stochastic and can hallucinate.

\subsection{Stage 3: Manual Expert Review and Final Selection}

The top-ranked candidates from Stage 2 were subjected to manual expert review to apply domain-specific judgment. The exclusion criteria focused on suitability for scientific symbolic discovery:

\begin{enumerate}[label=\alph*)]
    \item \textbf{Synthetic Data:} Datasets known or suspected to be generated from deterministic analytical expressions were excluded, avoiding tasks that reduce to reverse-engineering rather than genuine discovery.
    
    \item \textbf{Phenomena Lacking Stable Scientific Basis:} Datasets describing highly subjective or stochastic processes (e.g., human decision-making in games or social behaviors) were removed, as these are unlikely to yield universal symbolic laws.
\end{enumerate}

This process ensured diversity across domains (e.g., biology, astronomy, fluid dynamics) and relevance to observational phenomena. The final selection comprised 8 datasets, summarized in Tab.~\ref{tab:pmlb_datasets_app} (consistent with the overview in the main text). These datasets emphasize noisy, real-world variability without ground-truth equations, providing a stringent testbed for accuracy-complexity trade-offs.

\begin{table}[ht]
\centering
\caption{\textbf{Overview of eight real-world PMLB datasets for S without ground-truth equations.} Curated via three-stage pipeline (Appendix~\ref{app:dataset_selection}); columns: points ($N$), features, description. Tests accuracy-complexity trade-offs under noise (cf. Fig.~\ref{fig:NMSE}).}
\label{tab:pmlb_datasets_app}
\begin{tabular}{@{}lccc>{\centering\arraybackslash}p{0.4\linewidth}@{}}
\toprule
Dataset & Points& Features& & Description \\ \midrule
192\_vineyard & 52 & 2 & & Yields of grape harvests in vineyard rows \\
first\_principles\_supernovae\_zr & 236 & 1 & & Normalized flux of a supernova explosion \\
first\_principles\_supernovae\_zg & 243 & 1 & & Normalized flux of a supernova explosion \\
560\_bodyfat & 252 & 14 & & Body fat percentage estimates for men \\
nikuradse\_2 & 362 & 1 & & Resistance for fluid flow in rough pipes \\
nikuradse\_1 & 362 & 2 & & Resistance for fluid flow in rough pipes \\
solar\_flare & 1066 & 10 & & Number of minor C-class solar flares \\
529\_pollen & 3848 & 4 & & Pollen mass and geometry factors \\
\bottomrule
\end{tabular}
\end{table}

\section{Top Models from Symbolic Regression Methods on PMLB Datasets}
\label{app:top_models}

To illustrate comparative behavior of IdeaSearchFitter and baselines on the curated PMLB datasets (Appendix~\ref{app:dataset_selection}), this appendix presents the top-3 candidate models generated by each method. For each dataset, models are ranked by increasing validation NMSE, drawn from the Pareto frontier of solutions (Section~\ref{ssec:pmlb}). Equations are reported post-simplification, with parameters fitted via least-squares optimization on the training split (75\%-25\% train-validation). 

IdeaSearchFitter's models emphasize semantic coherence and interpretability, often incorporating domain-motivated structures. In contrast, baselines like Operon and PySR frequently produce more complex or oscillatory forms, 
highlighting trade-offs in the accuracy-interpretability plane. AI-Feynman failures are noted where applicable, in our experiments these failures commonly coincided with small sample sizes or higher input dimensionality. 

\paragraph{Caveat:} These observations are empirical and conditioned on our implementation choices and dataset selection.

The results are summarized in Tab.~\ref{tab:appendix_formulas}.

\begin{longtable}{llp{0.6\textwidth}r}
\caption{Top-3 models from each S method on the eight PMLB datasets, ranked by validation NMSE. Equations are in canonical form with fitted parameters; complexity (node count) is omitted for brevity but aligns with Pareto selection criteria (Section~\ref{ssec:pmlb}). Failures for AI-Feynman are indicated.}
\label{tab:appendix_formulas} \\
\toprule
\textbf{Method} & \textbf{Rank} & \textbf{Fitted Equation} & \textbf{Val. NMSE} \\
\midrule
\endfirsthead
\multicolumn{4}{c}%
{{\bfseries \tablename\ \thetable{} -- continued from previous page}} \\
\toprule
\textbf{Method} & \textbf{Rank} & \textbf{Fitted Equation} & \textbf{Val. NMSE} \\
\midrule
\endhead
\bottomrule
\endfoot
\multicolumn{4}{l}{\textbf{192 vineyard}} \\
\midrule
IdeaSearch & 1 & \texttt{6.9322*log(9*lugs\_1989/2 + lugs\_1990) - 3.2681} & 0.1133 \\
IdeaSearch & 2 & \texttt{6.1426*log(4.5946*lugs\_1989 + lugs\_1990) - 0.8782} & 0.1287 \\
IdeaSearch & 3 & \texttt{2.9976*log(lugs\_1989*lugs\_1990 + 1) + 2.9976*tanh((lugs\_1989 - lugs\_1990)/(lugs\_1989 + lugs\_1990)) + 9.9030} & 0.1727 \\
\midrule
Operon & 1 & \texttt{(((((59.0423 + 54.2717) * (((30.9482 + ((-27.3178) + ((-0.4296) * lugs\_1990))) / (65.5491 + ((-14.8242) * lugs\_1989))) / ((739.7688 * lugs\_1990) + ((-7804.3013) * lugs\_1989)))) + (0.0038 + (0.0028 * lugs\_1989))) * (2161.9905 + ((-21.5387) + ((-190.8566) * lugs\_1989)))) + ((((-0.1185) * lugs\_1990) + (1.3674 * lugs\_1989)) / (((-110.8519) + (8.5815 * lugs\_1990)) + (1.7410 * lugs\_1989))))} & 0.5846 \\
\midrule
PySR & 1 & \texttt{((14.1401 - lugs\_1990) * (sqrt(lugs\_1989) + -1.8355)) + 19.6876} & 0.1600 \\
PySR & 2 & \texttt{(-46.5294 / (lugs\_1989 + sqrt(lugs\_1990))) - -26.5472} & 0.1803 \\
PySR & 3 & \texttt{(-39.1905 / (lugs\_1989 + 2.1609)) - -26.4285} & 0.1889 \\
\midrule
AI-Feynman & 1 & \texttt{Failed} & NaN \\
\midrule
\multicolumn{4}{l}{\textbf{529 pollen}} \\
\midrule
IdeaSearch & 1 & \texttt{(-0.8200*WEIGHT*(Abs(CRACK) + Abs(NUB) + Abs(RIDGE) + 1) + 5.3977*(NUB - RIDGE)*(Abs(CRACK*NUB*RIDGE) + 1.7662))/((Abs(CRACK*NUB*RIDGE) + 1.7662)*(Abs(CRACK) + Abs(NUB) + Abs(RIDGE) + 1))} & 0.5227 \\
IdeaSearch & 2 & \texttt{0.0530 * (WEIGHT - (RIDGE + NUB + CRACK))} & 0.8985 \\
IdeaSearch & 3 & \texttt{-0.0568*CRACK - 0.0568*NUB - 0.0568*RIDGE + 0.0484*WEIGHT} & 0.8987 \\
\midrule
Operon & 1 & \texttt{((((-0.2133) * CRACK) + 0.4726) + (((-1.2542) / ((49.2894 * CRACK) + (-34.9567))) + (((((((-0.0882) * RIDGE) + (0.0231 * NUB)) + ((-0.0391) * WEIGHT)) + ((0.0005 / (((-0.8679) * CRACK) + 0.5654)) + ((-0.0511) + (0.0001 * NUB)))) + (0.0196 * CRACK)) * ((((-0.5127) + ((0.0177 * NUB) * (8.5303 + (2.3175 * WEIGHT)))) / ((4.5543 * RIDGE) + 4.2133)) + (((0.0231 * WEIGHT) + (9.5566 + ((-0.0355) * CRACK))) + ((0.0562 * NUB) + ((-0.0062) * RIDGE)))))))} & 0.2389 \\
\midrule
PySR & 1 & \texttt{(NUB - (((RIDGE + (WEIGHT * 0.4527)) / 0.2503) + ((CRACK * 0.1208) - sin((RIDGE / -1.5419) - -5852925.5000)))) * 0.2099} & 0.1995 \\
PySR & 2 & \texttt{(((NUB + sin(RIDGE * -0.6650)) - (((WEIGHT * 0.4525) + RIDGE) / 0.2497)) * 0.2095) + (CRACK * -0.0254)} & 0.1999 \\
PySR & 3 & \texttt{((CRACK * -0.1223) + ((NUB + sin((NUB / ((CRACK + WEIGHT) + 11.9499)) - (RIDGE * 0.6994))) - (((WEIGHT * 0.4524) + RIDGE) / 0.2522))) * 0.2118} & 0.2006 \\
\midrule
AI-Feynman & 1 & \texttt{Failed} & NaN \\
\midrule
\multicolumn{4}{l}{\textbf{560 bodyfat}} \\
\midrule
IdeaSearch & 1 & \texttt{0.0481*log(Abdomen/Wrist) - 450 + 495/Density} & 0.0001 \\
IdeaSearch & 2 & \texttt{(Density*(0.0423*log((Abdomen + Chest + Hip)/(Ankle + Wrist)) - 450) + 495)/Density} & 0.0001 \\
IdeaSearch & 3 & \texttt{0.0758*log10(Abdomen/Neck) + 0.2876*tanh(Age/20 - 2) - 450 + 495/Density} & 0.0004 \\
\midrule
Operon & 1 & \texttt{(((((2.5581 / 48.3727) / ((((-3.3820) * Density4) + ((-453.9575) * Density)) + (484.5315 + (1.1399 * Density2)))) + (511.6945 * (4.5171 / (4.5913 * Density)))) + ((((((4.4445 / (4.3245 * Density)) / (-33.3423)) / ((((-24.0674) * Density4) + ((-400.5348) * Density)) + (385.4976 + (16.4553 * Density2)))) + ((-1429.1804) + (507.4639 + (((-0.0062) * Hip) + (-521.5176))))) + 978.6653) + ((0.0225 * Density4) + (6.4951 * Density)))) + ((0.0003 * Chest) + ((-0.0033) * Density3)))} & 0.0000 \\
\midrule
PySR & 1 & \texttt{(-433.3768 + ((square(((0.2736 / cos((Age3 * 10.9942) + Neck)) - (2.6259 / cos((Age3 / -1.6016) + -1.3810))) / Age1) + (476.7212 / Density)) - (0.0037 / sin(Age0 / ((Abdomen - Thigh) / Neck))))) / ((square(0.1802 / sin(Neck + Age)) / 188696.8100) + 0.9640)} & 0.0000 \\
PySR & 2 & \texttt{((476.7212 / Density) + ((square(((0.2723 / cos(Neck + (Age3 * 10.9942))) - (2.6239 / cos(-1.3810 + (Age3 / -1.6016)))) / Age1) - (0.0037 / sin(Chest / -2.5985))) + -433.3691)) / 0.9648} & 0.0000 \\
PySR & 3 & \texttt{(((476.7212 / Density) + -433.3768) + (square(((2.6239 / cos((Age3 / -1.6016) + -1.3810)) - (0.2737 / cos(Neck + (Age3 * 10.9942)))) / Age1) - (0.0037 / sin(Age0 / ((Abdomen - Thigh) / Neck))))) / (((0.0415 / sin(Neck + Age)) / 434.3962) + 0.9641)} & 0.0000 \\
\midrule
AI-Feynman & 1 & \texttt{Failed} & NaN \\
\midrule
\multicolumn{4}{l}{\textbf{first principles supernovae zr}} \\
\midrule
IdeaSearch & 1 & \texttt{exp(-0.0215*Xaxis0 - 0.0022*(Xaxis0 - Abs(Xaxis0))**2 - 0.0215*Abs(Xaxis0))} & 0.0144 \\
IdeaSearch & 2 & \texttt{exp((0.0298*arctan(Xaxis0) - 0.0899)*log(cosh(Xaxis0)))} & 0.0156 \\
IdeaSearch & 3 & \texttt{exp(-Xaxis0*(0.0844*tanh(Xaxis0) - 0.0413))} & 0.0163 \\
\midrule
Operon & 1 & \texttt{(((((((0.0290 * Xaxis0) + (-33.9033)) + (((-0.0019) * Xaxis0) + 34.7889)) + ((0.0026 * Xaxis0) * (((-0.2686) * Xaxis0) + (-0.5637)))) + ((((0.0003 * Xaxis0) * ((-1.6971) * Xaxis0)) + (0.1610 * Xaxis0)) * (((-0.0070) * Xaxis0) * ((-0.0060) * Xaxis0)))) + (((0.0047 * Xaxis0) + (-0.1858)) / (((0.0901 * Xaxis0) + (-1.7062)) + ((0.0012 * Xaxis0) * ((-1.2067) * Xaxis0))))) + (((((-0.0500) * Xaxis0) * ((-1.6661) * Xaxis0)) + (0.3720 * Xaxis0)) / (((0.5639 * Xaxis0) + (-15.1843)) + ((0.0230 * Xaxis0) * ((-4.2815) * Xaxis0)))))} & 0.0058 \\
\midrule
PySR & 1 & \texttt{0.9688 / sqrt(exp((Xaxis0 + cos(Xaxis0)) * 0.0834) + (0.0000 / exp(Xaxis0)))} & 0.0146 \\
PySR & 2 & \texttt{sqrt((0.9305 / ((0.0000 / exp(Xaxis0)) + exp(Xaxis0 * 0.0830))) - 0.0004)} & 0.0152 \\
PySR & 3 & \texttt{sqrt(0.9341 / ((0.0000 / exp(Xaxis0)) + exp(Xaxis0 * 0.0835)))} & 0.0161 \\
\midrule
AI-Feynman & 1 & \texttt{0.0469*pi} & 1.0880 \\
\midrule
\multicolumn{4}{l}{\textbf{first principles supernovae zg}} \\
\midrule
IdeaSearch & 1 & \texttt{2.5695*exp(-0.1016*Xaxis0 - 0.9453*exp(-0.1016*Xaxis0))} & 0.0037 \\
IdeaSearch & 2 & \texttt{exp(-0.0988*Xaxis0 + 1 - exp(-0.0988*Xaxis0))} & 0.0074 \\
IdeaSearch & 3 & \texttt{(0.0423*Xaxis0 + 1)**5.1426*exp(-0.2176*Xaxis0)} & 0.0087 \\
\midrule
Operon & 1 & \texttt{(((0.0055 * Xaxis0) + (-0.5684)) * (((((((-0.0094) * Xaxis0) + (-0.2051)) + ((((0.0138 * Xaxis0) + (-0.8186)) * (0.0099 + (0.0005 * Xaxis0))) * ((7.8337 + ((-0.3604) * Xaxis0)) * ((0.1445 * Xaxis0) + 0.8769)))) * (((((-0.0145) * Xaxis0) + 1.1968) * (4.1841 + ((-0.0740) * Xaxis0))) * (8.3004 + ((-0.0949) * Xaxis0)))) + ((-0.0891) * Xaxis0)) / ((((-1.5215) + (0.5994 * Xaxis0)) * ((0.0117 * Xaxis0) + 0.3677)) + (7.4031 + (-0.6261)))))} & 0.0003 \\
\midrule
PySR & 1 & \texttt{((-0.0046 / (sin(Xaxis0 + -1.1818) - (Xaxis0 / -6.7722))) - (square(square(sin(Xaxis0 * 0.0263))) * -0.0225)) + square((sin(Xaxis0 * 0.1614) * 0.0126) + square(square(sin(0.2667 - (11.4245 / ((Xaxis0 * 0.1219) + 6.3517))))))} & 0.0002 \\
PySR & 2 & \texttt{square((square(square(sin((11.4245 / ((0.1222 * Xaxis0) + 6.3517)) - 0.2667))) / 0.9986) + (sin(Xaxis0 * 0.1614) * 0.0126)) + ((-0.0046 / (sin(Xaxis0 + -1.1818) - (Xaxis0 * -0.1458))) - (square(square(sin(Xaxis0 * 0.0264))) * -0.0225))} & 0.0002 \\
PySR & 3 & \texttt{(((-0.0022 / (sqrt(square(Xaxis0)) - 4.8672)) + 1.0025) * square((sin(Xaxis0 * 0.1609) * 0.0120) + square(square(sin(0.2683 - (11.4240 / ((Xaxis0 * 0.1220) + 6.3522))))))) - (square(square(sin(Xaxis0 * 0.0264))) * -0.0223)} & 0.0002 \\
\midrule
AI-Feynman & 1 & \texttt{0.4747*exp((Xaxis0/(-exp(pi))))} & 0.6974 \\
AI-Feynman & 2 & \texttt{0.0002+(pi/(Xaxis0-1))} & 1.2573 \\
AI-Feynman & 3 & \texttt{acos(-666.0000*sin(pi))} & 27.5223 \\
\midrule
\multicolumn{4}{l}{\textbf{nikuradse 2}} \\
\midrule
IdeaSearch & 1 & \texttt{-0.3888*log(8.7331*log\_v\_k\_nu + 1) + 2.9607*tanh(2.1826*log\_v\_k\_nu)} & 0.1880 \\
IdeaSearch & 2 & \texttt{1.6562 + 1.5284*exp(-log\_v\_k\_nu) - 2.4917*exp(-2*log\_v\_k\_nu)} & 0.2372 \\
IdeaSearch & 3 & \texttt{-0.2225*log\_v\_k\_nu - 1.5372*arctan(-5.7640*log\_v\_k\_nu)} & 0.3250 \\
\midrule
Operon & 1 & \texttt{(((0.1015 / (126.2199 + ((-174.7875) * log\_v\_k\_nu))) + ((((-14.7983) * log\_v\_k\_nu) + (0.4857 + (16.0479 * log\_v\_k\_nu))) + (((((133.6521 + ((-174.5841) * log\_v\_k\_nu)) / (11.3310 + ((-14.7983) * log\_v\_k\_nu))) + (97.6692 * log\_v\_k\_nu)) + ((9.5892 * log\_v\_k\_nu) * ((-7.4763) * log\_v\_k\_nu))) / ((((-130.7082) * log\_v\_k\_nu) + 120.6669) + (((-8.1544) * log\_v\_k\_nu) * ((-9.2046) * log\_v\_k\_nu)))))) + (((0.0127 * log\_v\_k\_nu) / (28.7919 + ((-45.8219) * log\_v\_k\_nu))) + (((0.0295 * log\_v\_k\_nu) * ((-6.2251) * log\_v\_k\_nu)) + ((0.0109 * log\_v\_k\_nu) / (((-95.3552) * log\_v\_k\_nu) + 100.2356)))))} & 0.0317 \\
\midrule
PySR & 1 & \texttt{square((sin(log\_v\_k\_nu + -0.5354) * exp(square(log\_v\_k\_nu) * -1.1018)) + 1.3121)} & 0.0235 \\
PySR & 2 & \texttt{square(((-0.5117 + sin(log\_v\_k\_nu)) * exp(log\_v\_k\_nu * (0.1898 - log\_v\_k\_nu))) + 1.3133)} & 0.0238 \\
PySR & 3 & \texttt{square(((log\_v\_k\_nu + -0.5459) * exp((sqrt(log\_v\_k\_nu) / -1.2050) * (square(log\_v\_k\_nu) + 0.3561))) + 1.3140)} & 0.0254 \\
\midrule
AI-Feynman & 1 & \texttt{1.0640+sin((sqrt(log\_v\_k\_nu)-cos((log\_v\_k\_nu+1))))} & 0.1102 \\
AI-Feynman & 2 & \texttt{1.0108+cos(((pi/exp(log\_v\_k\_nu))-1))} & 0.1419 \\
AI-Feynman & 3 & \texttt{1.5523+(log\_v\_k\_nu/exp(log\_v\_k\_nu))} & 0.7225 \\
\midrule
\multicolumn{4}{l}{\textbf{nikuradse 1}} \\
\midrule
IdeaSearch & 1 & \texttt{-(log\_Re/8 - 3/4)*(tanh(-2.9607*log\_Re + 2.9607*log10(2*r\_k) + 6.4621) + 1) - (log10(25)/2 - log10(log10(37*r\_k/5)))*(tanh(-2.9607*log\_Re + 2.9607*log10(2*r\_k) + 6.4621) - 1)} & 0.0030 \\
IdeaSearch & 2 & \texttt{-(log\_Re - 6)*(tanh(-6.5881*log\_Re + 6.5881*log10(2*r\_k) + 15.4972) + 1)/8 + (log10(2*log10(7.2341*r\_k)) - 1)*(tanh(-6.5881*log\_Re + 6.5881*log10(2*r\_k) + 15.4972) - 1)} & 0.0132 \\
IdeaSearch & 3 & \texttt{-(log\_Re - 6)*(tanh(-4.7452*log\_Re + 4.7452*log10(r\_k) + 8.0275) + 1)/8 + (log10(2*log10(7.9565*r\_k)) - 1)*(tanh(-4.7452*log\_Re + 4.7452*log10(r\_k) + 8.0275) - 1)} & 0.0311 \\
\midrule
Operon & 1 & \texttt{(((((((0.4479 * log\_Re) + (-0.9317)) + (((-0.0002) + ((-0.0001) * r\_k)) / ((-16.6700) + (3.9010 * log\_Re)))) + (((-0.1017) + (0.0099 * r\_k)) / ((-7.8842) + (2.4935 * log\_Re)))) + ((0.0014 + ((-0.0000) * r\_k)) / ((-16.7653) + (3.8393 * log\_Re)))) + (((((-0.9019) * log\_Re) * (0.0394 * log\_Re)) + (0.0016 * r\_k)) + ((((-0.0003) * r\_k) + 2.0806) * ((0.1261 * r\_k) / ((-11.8721) * log\_Re))))) + (((-8.0012) + (0.7983 * log\_Re)) / ((-5.6613) + ((-0.2989) * r\_k))))} & 0.2310 \\
\midrule
PySR & 1 & \texttt{(0.1532 * sin((square(log(r\_k)) + r\_k) / exp(log\_Re * log(log\_Re)))) + sqrt(sqrt(1.7694 / ((1.6665 + (r\_k - (cos(log\_Re) * 0.4594))) * (square(square((cos((log(sqrt(r\_k)) - log\_Re) / -0.6659) * -0.4567) + 0.2502)) + 0.2706))) + -0.0235)} & 0.0014 \\
PySR & 2 & \texttt{(sin((r\_k + square(log(r\_k))) / exp(log(log\_Re) * log\_Re)) * 0.1532) + sqrt(sqrt(1.7693 / (((r\_k - (cos(square(r\_k + 0.4175)) * cos(log\_Re))) + 1.6493) * (square(square((cos((log\_Re - log(sqrt(r\_k))) / -0.6652) * -0.4563) + 0.2496)) + 0.2714))) + -0.0234)} & 0.0014 \\
PySR & 3 & \texttt{sqrt(sqrt(1.7706 / ((r\_k + 1.7035) * (square(square(0.2438 + (cos((log(sqrt(0.5224 + r\_k)) - log\_Re) / 0.6684) * -0.4481))) + 0.2660))) + -0.0248) + (sin(r\_k / (exp(log\_Re * log(log\_Re)) - square(log(r\_k)))) * 0.1542)} & 0.0014 \\
\midrule
AI-Feynman & 1 & \texttt{atan(0.1703+(pi/sqrt(r\_k)))} & 0.0687 \\
\midrule
\multicolumn{4}{l}{\textbf{solar flare}} \\
\midrule
IdeaSearch & 1 & \texttt{1.5284*tanh(0.0174*sunspot\_distribution*(sunspot \_group\_configuration + sunspot\_largest\_spot)*(historically \_complex\_region\_recent + recent\_flare\_activity + sunspot\_activity + sunspot\_evolution))} & 0.7892 \\
IdeaSearch & 2 & \texttt{1.0175*tanh(sqrt(region\_area)* 
(historically\_ complex\_region\_recent + recent\_flare\_activity) + 0.1838*sunspot\_distribution*sunspot\_group\_configuration)} & 0.8006 \\
IdeaSearch & 3 & \texttt{0.0317*sunspot\_group\_configuration*(historically \_complex\_region\_recent + 1)*(sunspot \_activity + sunspot\_evolution + 1)*arctan(sunspot\_distribution*sunspot\_largest\_spot) *exp(0.3437*recent\_flare\_activity)} & 0.8501 \\
\midrule
Operon & 1 & \texttt{((((((-0.2405) * recent\_flare\_activity) + (0.2552 * sunspot\_distribution)) * ((-0.9512) * sunspot\_group\_configuration)) * ((-0.9512) * sunspot\_group\_configuration)) * (((((-0.4275) * sunspot\_group\_configuration) + (((1.0932 * sunspot\_activity) * ((-0.9512) * sunspot\_group\_configuration)) + (2.5258 * sunspot\_activity))) + ((((((-0.2405) * recent\_flare\_activity) + (0.2552 * sunspot\_distribution)) * (1.5155 * historically\_complex\_region)) + (1.0932 * sunspot\_activity)) + ((((0.6356 * sunspot\_largest\_spot) + (1.0932 * sunspot\_activity)) + ((1.0932 * sunspot\_activity) * ((-0.9512) * sunspot\_group\_configuration))) + ((0.2552 * sunspot\_distribution) + ((-0.2405) * recent\_flare\_activity))))) + ((((0.2552 * sunspot\_distribution) + (1.0932 * sunspot\_activity)) * ((1.1720 * sunspot\_evolution) * (1.5155 * historically\_complex\_region))) * (((-3.3306) * sunspot\_distribution) + 8.6404))))} & 191.9594 \\
\midrule
PySR & 1 & \texttt{exp(-1.0489 - cos(sunspot\_group\_configuration))} & 0.8156 \\
PySR & 2 & \texttt{exp((sunspot\_activity + -1.5812) - cos(sunspot\_group\_configuration))} & 0.8159 \\
PySR & 3 & \texttt{square(sin(sqrt(sunspot\_group\_configuration * 7.1457)))} & 0.8237 \\
\midrule
AI-Feynman & 1 & \texttt{Failed} & NaN \\
\midrule

\end{longtable}

\section{Extended Results for 2D PDF Fits and Extrapolation}
\label{app:full_2d}

This appendix reports the complete set of quantitative results and diagnostic figures referenced in Section~\ref{ssec:pdf}. All plots use the CT18NNLO data on $x\in[10^{-3},0.4]$ and training scales $Q\in{2,5}\,\mathrm{GeV}$, with out-of-domain validation at $Q=100\,\mathrm{GeV}$. Reduced $\chi^2$ is used throughout; model complexity is the canonical expression-tree node count. Figures are grouped by flavour.

\subsection{Training–validation chi2 Pareto fronts (combined)}
\label{app:combined_fronts}

Each panel compares IdeaSearchFitter and PySR on the same axes: training $\chi^2$ (x, log scale) and validation $\chi^2$ (y, log scale). Point colour encodes complexity.

\begin{figure}[htbp]
\centering
\begin{minipage}[t]{0.49\linewidth}
  \centering
  \includegraphics[width=\linewidth]{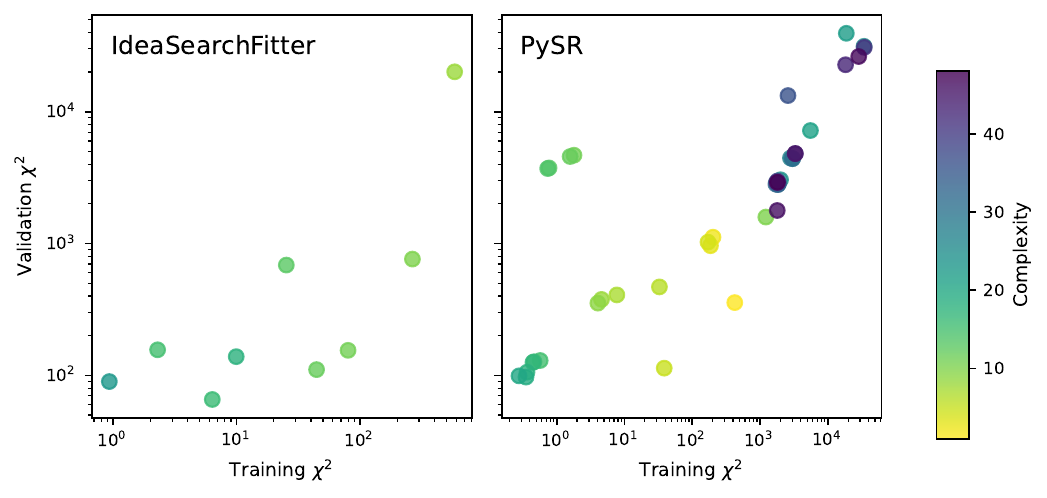}
  \caption{Combined Pareto fronts for $u_v$.}
  \label{fig:app_pareto_combined_uv}
\end{minipage}\hfill
\begin{minipage}[t]{0.49\linewidth}
  \centering
  \includegraphics[width=\linewidth]{fig/pdf/figure8_pareto_front_ubar.pdf}
  \caption{Combined Pareto fronts for $\bar{u}$.}
  \label{fig:app_pareto_combined_ubar}
\end{minipage}
\end{figure}

\begin{figure}[htbp]
\centering
\begin{minipage}[t]{0.49\linewidth}
  \centering
  \includegraphics[width=\linewidth]{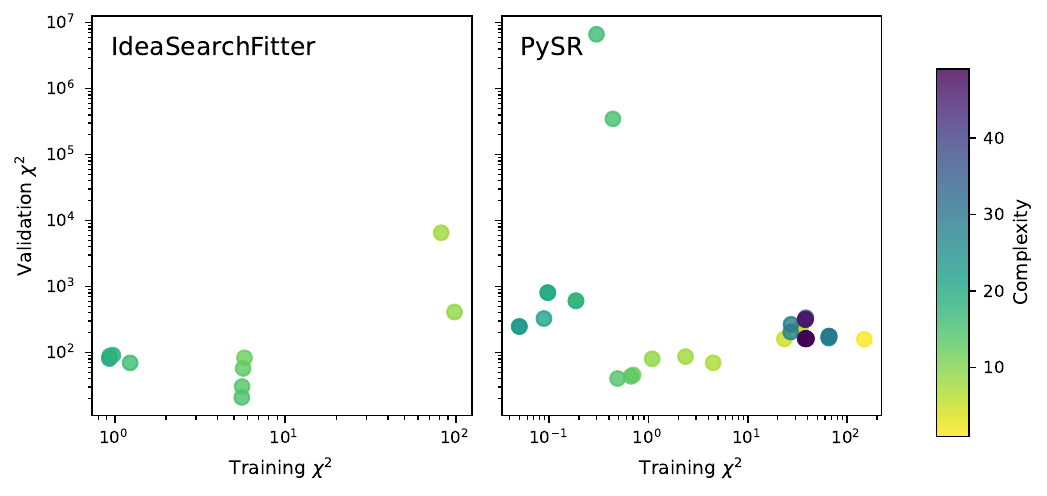}
  \caption{Combined Pareto fronts for $d_v$.}
  \label{fig:app_pareto_combined_dv}
\end{minipage}\hfill
\begin{minipage}[t]{0.49\linewidth}
  \centering
  \includegraphics[width=\linewidth]{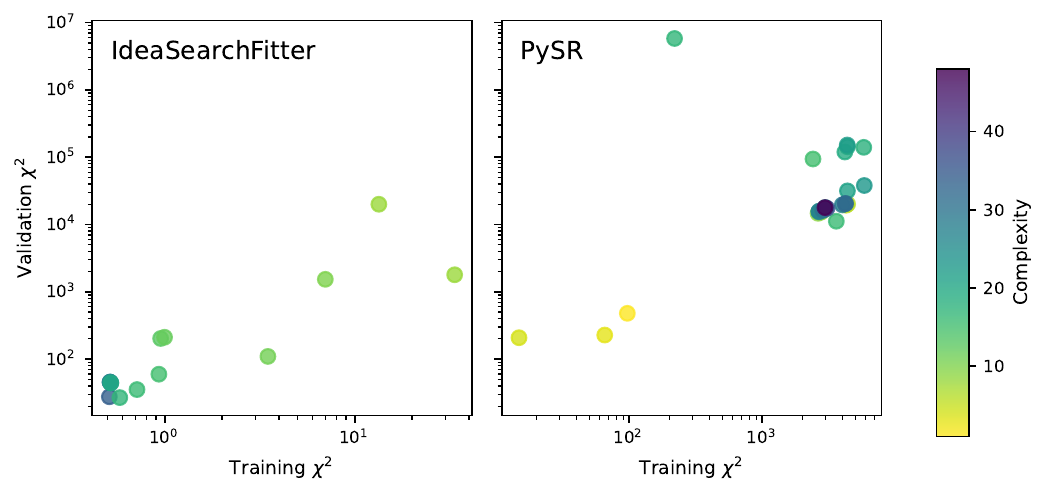}
  \caption{Combined Pareto fronts for $\bar{d}$.}
  \label{fig:app_pareto_combined_dbar}
\end{minipage}
\end{figure}

\begin{figure}[htbp]
\centering
\begin{minipage}[t]{0.49\linewidth}
  \centering
  \includegraphics[width=\linewidth]{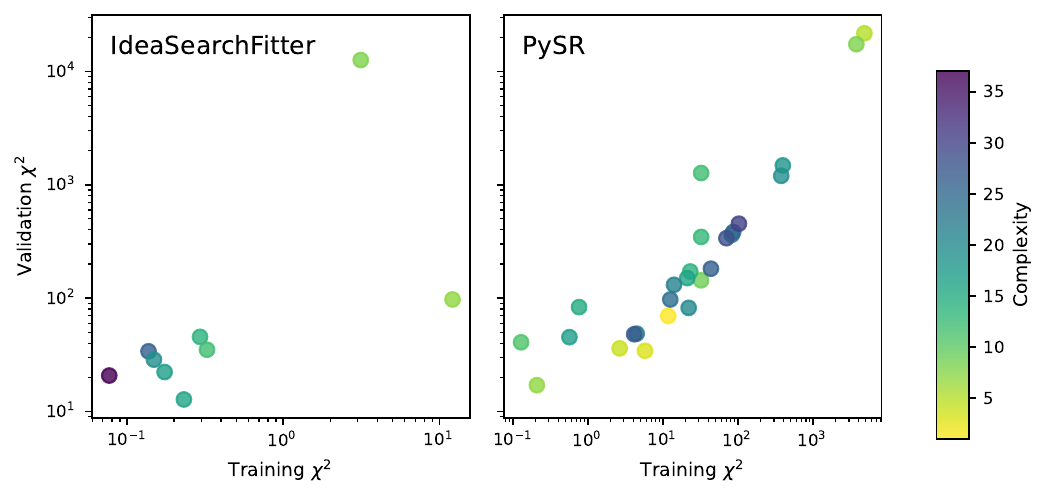}
  \caption{Combined Pareto fronts for $s$.}
  \label{fig:app_pareto_combined_s}
\end{minipage}\hfill
\begin{minipage}[t]{0.49\linewidth}
  \centering
  \includegraphics[width=\linewidth]{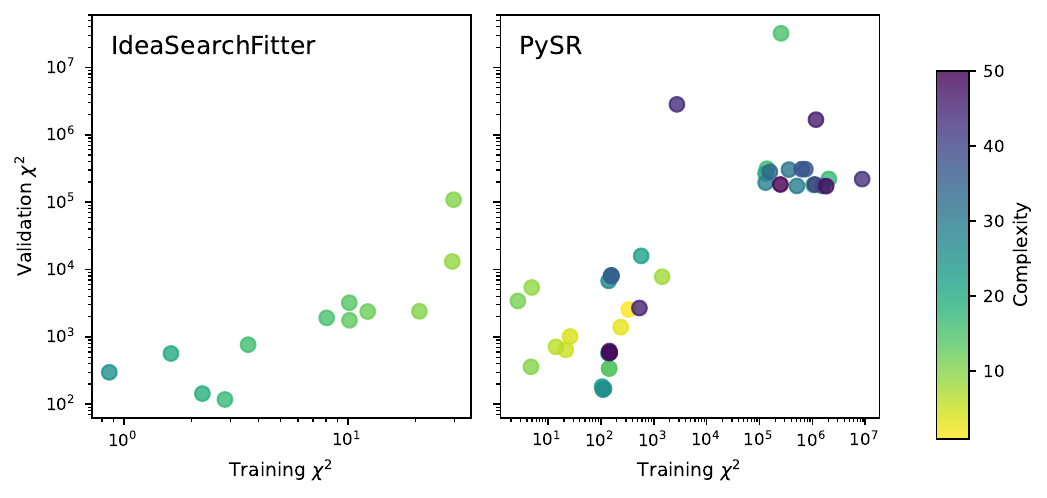}
  \caption{Combined Pareto fronts for $g$.}
  \label{fig:app_pareto_combined_g}
\end{minipage}
\end{figure}

\subsection{Scale-resolved comparisons: in-domain (Q=2,5 GeV) vs.\ out-of-domain (Q=100 GeV)}
\label{app:scale_resolved}

For each flavour, we show the best-validation models from both methods at $Q=2$ and $Q=5\,\mathrm{GeV}$ (training scales) and at $Q=100\,\mathrm{GeV}$ (extrapolation). Blue points: reference grids; shaded bands: reference uncertainty; red dashed: IdeaSearchFitter; green dotted: PySR.

\begin{figure}[ht!]
\centering
\begin{subfigure}[b]{.32\linewidth}
\includegraphics[width=\linewidth]{fig/pdf/figure9_best_fit_ubar_train_q2.pdf}
\caption{$\bar{u}$ at $Q=2\,\mathrm{GeV}$}
\end{subfigure}\hfill
\begin{subfigure}[b]{.32\linewidth}
\includegraphics[width=\linewidth]{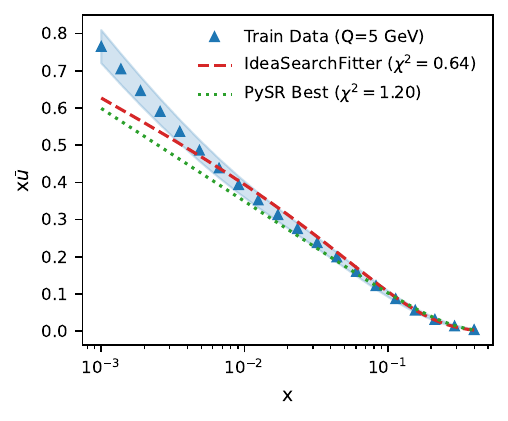}
\caption{$\bar{u}$ at $Q=5\,\mathrm{GeV}$}
\end{subfigure}\hfill
\begin{subfigure}[b]{.32\linewidth}
\includegraphics[width=\linewidth]{fig/pdf/figure9_best_fit_ubar.pdf}
\caption{$\bar{u}$ at $Q=100\,\mathrm{GeV}$}
\end{subfigure}
\label{fig:app_best_ubar}
\end{figure}

\begin{figure}[ht!]
\centering
\begin{subfigure}[b]{.32\linewidth}
\includegraphics[width=\linewidth]{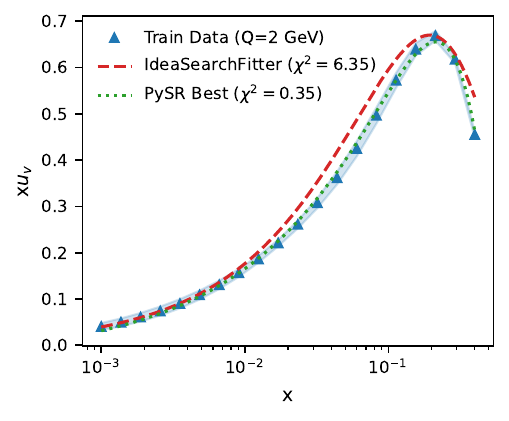}
\caption{$u_v$ at $Q=2\,\mathrm{GeV}$}
\end{subfigure}\hfill
\begin{subfigure}[b]{.32\linewidth}
\includegraphics[width=\linewidth]{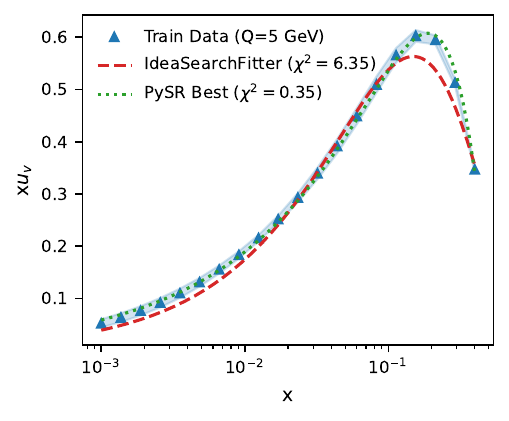}
\caption{$u_v$ at $Q=5\,\mathrm{GeV}$}
\end{subfigure}\hfill
\begin{subfigure}[b]{.32\linewidth}
\includegraphics[width=\linewidth]{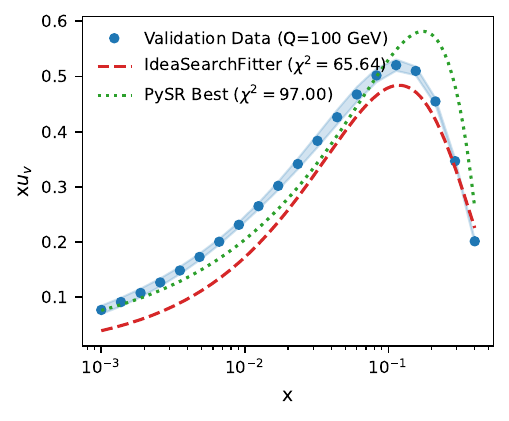}
\caption{$u_v$ at $Q=100\,\mathrm{GeV}$}
\end{subfigure}
\label{fig:app_best_uv}
\end{figure}

\begin{figure}[ht!]
\centering
\begin{subfigure}[b]{.32\linewidth}
\includegraphics[width=\linewidth]{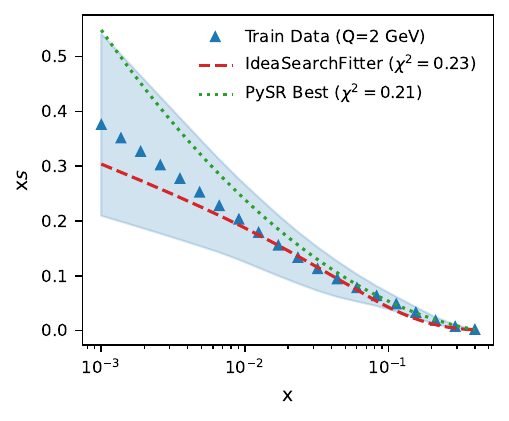}
\caption{$s$ at $Q=2\,\mathrm{GeV}$}
\end{subfigure}\hfill
\begin{subfigure}[b]{.32\linewidth}
\includegraphics[width=\linewidth]{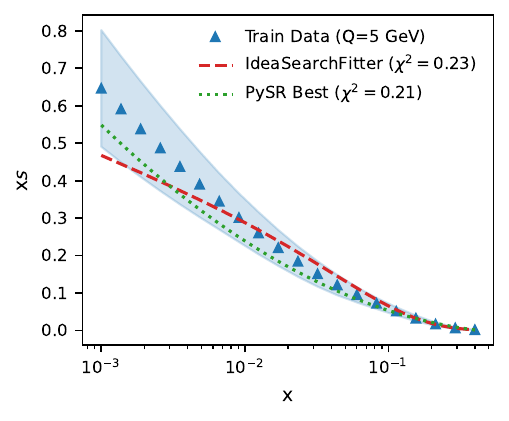}
\caption{$s$ at $Q=5\,\mathrm{GeV}$}
\end{subfigure}\hfill
\begin{subfigure}[b]{.32\linewidth}
\includegraphics[width=\linewidth]{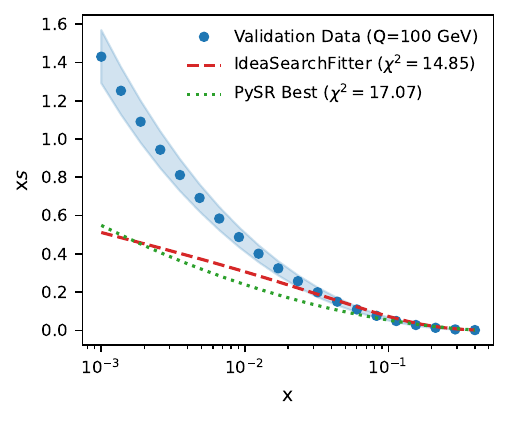}
\caption{$s$ at $Q=100\,\mathrm{GeV}$}
\end{subfigure}
\label{fig:app_best_s}
\end{figure}

\begin{figure}[ht!]
\centering
\begin{subfigure}[b]{.32\linewidth}
\includegraphics[width=\linewidth]{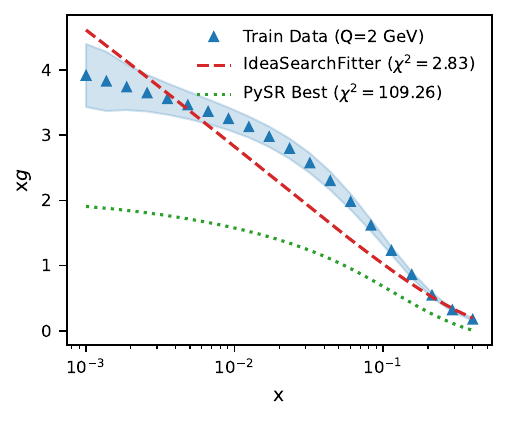}
\caption{$g$ at $Q=2\,\mathrm{GeV}$}
\end{subfigure}\hfill
\begin{subfigure}[b]{.32\linewidth}
\includegraphics[width=\linewidth]{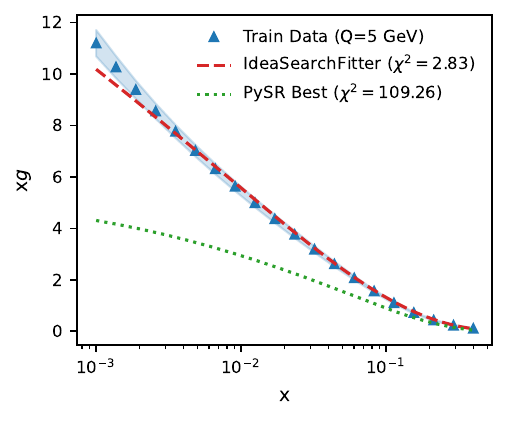}
\caption{$g$ at $Q=5\,\mathrm{GeV}$}
\end{subfigure}\hfill
\begin{subfigure}[b]{.32\linewidth}
\includegraphics[width=\linewidth]{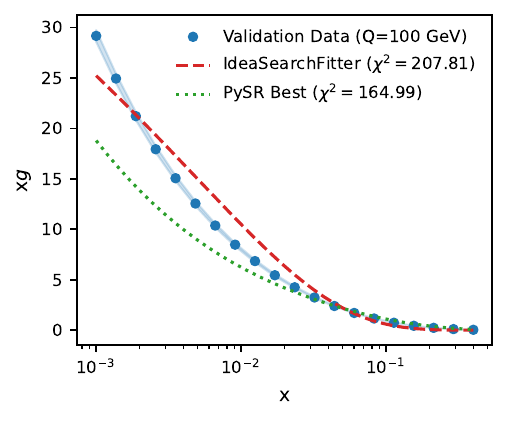}
\caption{$g$ at $Q=100\,\mathrm{GeV}$}
\end{subfigure}
\label{fig:app_best_g}
\end{figure}

\begin{figure}[ht!]
\centering
\begin{subfigure}[b]{.32\linewidth}
\includegraphics[width=\linewidth]{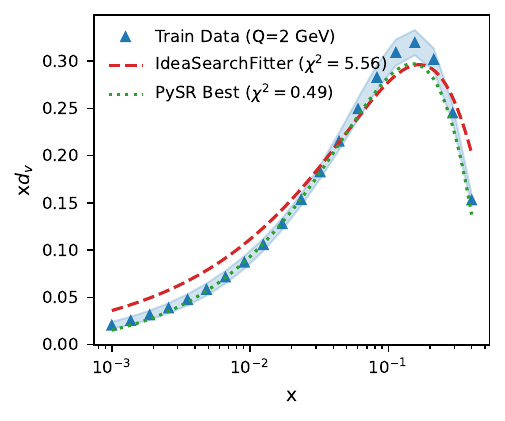}
\caption{$d_v$ at $Q=2\,\mathrm{GeV}$}
\end{subfigure}\hfill
\begin{subfigure}[b]{.32\linewidth}
\includegraphics[width=\linewidth]{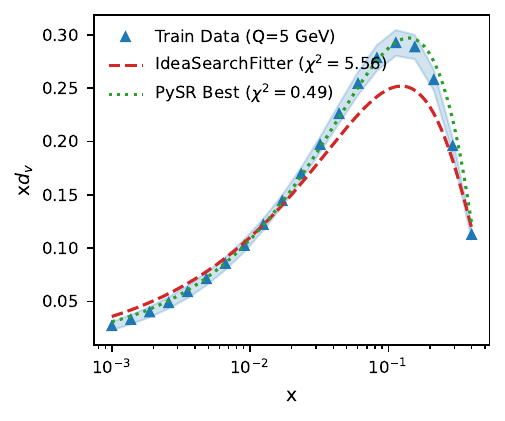}
\caption{$d_v$ at $Q=5\,\mathrm{GeV}$}
\end{subfigure}\hfill
\begin{subfigure}[b]{.32\linewidth}
\includegraphics[width=\linewidth]{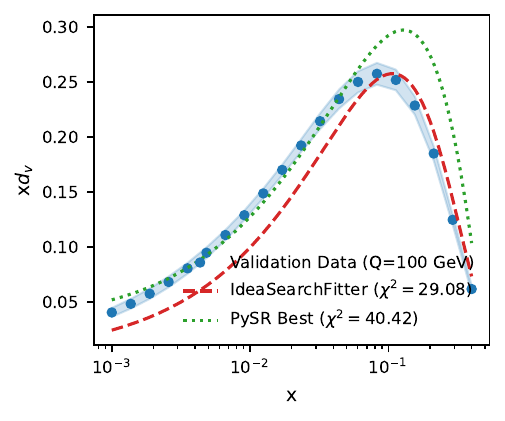}
\caption{$d_v$ at $Q=100\,\mathrm{GeV}$}
\end{subfigure}
\label{fig:app_best_dv}
\end{figure}

\begin{figure}[ht!]
\centering
\begin{subfigure}[b]{.32\linewidth}
\includegraphics[width=\linewidth]{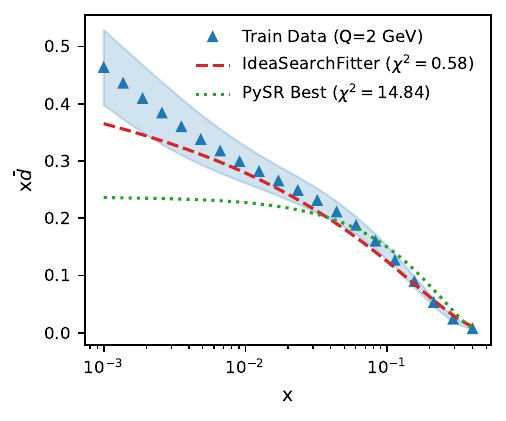}
\caption{$\bar{d}$ at $Q=2\,\mathrm{GeV}$}
\end{subfigure}\hfill
\begin{subfigure}[b]{.32\linewidth}
\includegraphics[width=\linewidth]{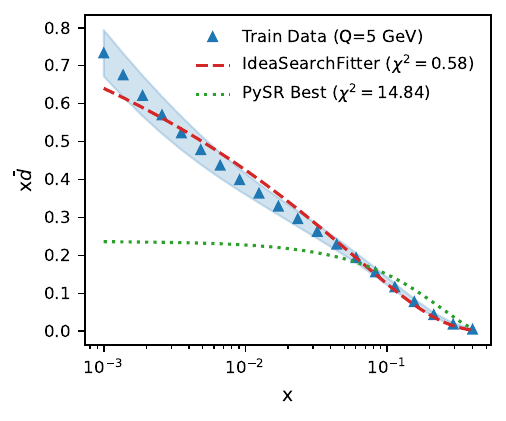}
\caption{$\bar{d}$ at $Q=5\,\mathrm{GeV}$}
\end{subfigure}\hfill
\begin{subfigure}[b]{.32\linewidth}
\includegraphics[width=\linewidth]{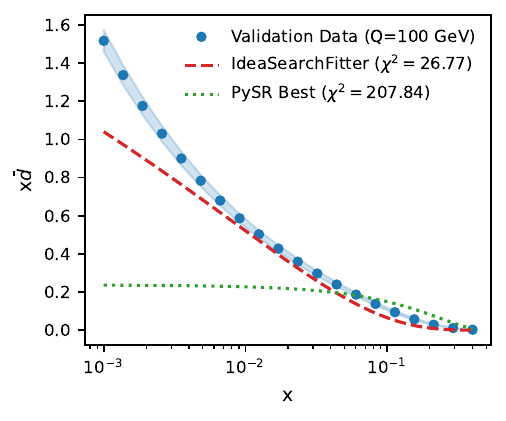}
\caption{$\bar{d}$ at $Q=100\,\mathrm{GeV}$}
\end{subfigure}
\label{fig:app_best_dbar}
\end{figure}

\end{document}